\documentclass{aa}
\usepackage{psfig}

\def\tmax{$t_{\rm max}$}
\def\kms{km/s}

\begin{document}

   \thesaurus{ 07 
		(03.13.2; 
		08.01.3;  
		08.12.2;  
		08.18.1;  
		08.19.6;  
		08.22.3   
	)
}

\title{Variability in ultra cool dwarfs: evidence for the evolution of surface features}
\titlerunning{Surface feature evolution in ultra cool dwarfs}

\subtitle{}

\author{C.A.L.\ Bailer-Jones \& R.\ Mundt}
\authorrunning{Bailer-Jones \& Mundt}
\offprints{\\Coryn Bailer-Jones, calj@mpia-hd.mpg.de}
\institute{Max-Planck-Institut f\"ur Astronomie, K\"onigstuhl 17, D-69117 Heidelberg, Germany}

\date{Received 17 October 2000; Accepted 1 December 2000}

\maketitle

\begin{abstract}

We present photometric light curves for a sample of 21 ultra cool M
and L dwarfs in the field and in the young open clusters
$\sigma$~Orionis and the Pleiades. The list of targets includes both
low mass hydrogen burning stars and brown dwarfs.  Evidence for
variability with RMS amplitudes (in the I band) of 0.01 to 0.055
magnitudes on timescales of 0.4 to 100 hours is discovered in half of
these objects.  Power spectral analysis using the {\sc CLEAN}
algorithm was performed to search for evidence of periodic
variability.  Some objects show strong periodicities at around a few
hours, which could be due to rotational modulation of the light curve
by surface features.  However, several objects do not have any
significant periodicities to explain their variability.  The $v \sin
i$ values of a similar population of objects makes it very likely that
our time sampling was sensitive to the expected range of rotation
periods, and simulations show that we would have detected these if
they were caused by long-lived surface features. We argue that this
absence of periodicity is due to the evolution of the brightness, and
presumably also the physical size, of surface features on timescales
of a few to a few tens of hours. This is supported in the case of
2M1145 for which two light curves have been obtained one year apart
and show no common periodicity. The surface features could plausibly
be photospheric dust clouds or magnetically-induced spots.  The
recently observed decline in chromospheric activity for late type M
and L dwarfs hints towards the former explanation for at least our
later-type objects. Furthermore, our sample suggests that variability
to be more common in objects later than M9, indicating that the
variability may be related to dust formation.  One light curve shows a
brief, but significant, dip, which could be a short-lived feature or
possibly an eclipse by a companion.


\end{abstract}

\section{Introduction}

Time-resolved observations are an important method for investigating
astrophysical phenomena. This is particularly the case for objects
which cannot be resolved spatially, as then the amount of information
available is greatly limited. Temporal monitoring is central to many
parts of astrophysics, such as pulsars, the physics of stars in the
instability strip, microlensing and gamma ray bursts.  However,
monitoring is important even for apparently ``stable'' objects, e.g.\
for the determination of stellar rotation periods, and has led to the
discovery of transient activity in a whole range of astrophysical
objects.

Variability is a phenomenon which is potentially important in ultra cool
dwarfs, because at these low temperatures (and masses) these objects
are fully convective, and many molecules and condensates form in their
atmospheres. Furthermore, many may also be rapid rotators, providing a
possible driving mechanism for atmospheric dynamics.  They could,
therefore, show a range of time-dependent observable phenomena,
such as the rotational modulation of the light curve due to surface
inhomogeneities, the evolution of magnetically-induced star spots,
accretion activity (for the youngest objects), flaring, movement of
photospheric clouds, and eclipses by unseen companions or disks.

Ultra cool dwarfs can be divided into the three spectral types T, L
and late M. The L dwarfs are the low temperature continuation of the M
dwarf sequence. As the temperature drops, the strong TiO and VO bands
which characterises the optical and infrared spectra of M dwarfs are
replaced by very broad neutral alkali lines and lines of iron
hydrides. Modelling of low resolution optical and near infrared
spectra implies a temperature range of 2000\,K down to 1300\,K
(Kirkpatrick et al.\ \cite{kirkpatrick00}). However, a temperature
assignment using fits to high resolution profiles of the alkali lines
indicates a somewhat hotter range of 2200--1600\,K (Basri et al.\
\cite{basrietal00}). At even lower temperatures methane can form, and
broad absorption features of this -- as well as water -- in the
infrared are the distinguishing features of T dwarfs. Ultra cool
dwarfs cover a range of masses (the mass for a given effective
temperature depending on the age) from a few Jupiter masses up to a
few tenths of a solar mass ($1 M_{\odot} = 1050 M_{\rm Jup}$).  For
example, an L dwarf could in principle, be a hydrogen burning star, a
brown dwarf or even a giant gas planet if it is young enough. Objects
later than L4.5 are expected to be substellar (Kirkpatrick et al.\
\cite{kirkpatrick00}).

To date, little variability monitoring of ultra cool dwarfs has been
reported. Tinney \& Tolley (\cite{tinney99}) found variability (at a
98\% confidence level) with an amplitude of 0.04 magnitudes over a few
hours in an M9 brown dwarf, but detected no variability above 0.1
magnitudes in an L5 dwarf.  Terndrup et al.\ (\cite{terndrup99})
searched for rotational modulation of the light curves of eight M type
stars and brown dwarfs in the Pleiades.  They derived periodicities
for two low mass stars, but found no significant variability in the
rest of the sample. At the lower end of the temperature scale,
Nakajima et al.\ (\cite{nakajima00}) found variability in the near
infrared spectrum of a T dwarf over a period of 80 minutes.  In an
earlier paper we reported the first results from a program to monitor
a number of brown dwarfs and L dwarfs (Bailer-Jones \&
Mundt~\cite{paper1}, hereafter paper I). Of the six objects monitored,
we discovered evidence for variability in the field L1 dwarf 2M1145,
and tentatively assigned a period. In the present paper we have
extended this work to a total of 21 M and L dwarfs, and look for
evidence of any variability in the I band down to a precision of 0.005
magniutdes on timescales between a fraction of an hour and several
days.

In the next section we describe the selection of the target objects
and their relevant properties. Section~\ref{data} describes the
observational and data reduction strategy, with a discussion of the
steps required to achieve high precision relative photometry on these
faint objects, as well as an accurate estimate of the photometric
errors. We then discuss the construction and analysis of the
differential light curves to look for evidence of variability.
Section~\ref{timeseries} describes our time series analysis
techniques. The results section summaries our findings, with a
description for individual objects. The main argument of this paper is
presented in section~\ref{discussion}, where we then discuss the
interpretation of our results in terms of physical phenomena. The data
presented in paper I have been re-reduced and re-analysed in the
present paper. Although the results are generally consistent, the
results in the present paper supersede those in paper I.

\section{Target selection}

\begin{table*}
\caption[]{Properties of ultra cool dwarf targets.  Each
reference makes use of a different I band and even definition of
magnitude, so values are only intended to be indicative. In particular,
the SDSS I filter is somewhat bluer than the Cousins I filter, thus yielding
fainter magnitudes for L dwarfs. The spectral types in parentheses have been
estimated from the R$-$I colours of B\'ejar et al.~(\cite{bejar99}).
\label{targets}
}
\begin{tabular}{lllllll}
\hline
name        	& IAU name               	& $I$	& SpT 	& H$\alpha$ EW	& $\ion{Li}{i}\,\lambda$6708 EW	& reference	\\
            	&                        	&	&   	& \AA\   	& \AA\                    	&		\\
\hline
2M0030		& \object{2MASSW J0030438$+$313932}	& 18.82	& L2	& 4.4 $\pm$0.2	& $< 1.0$			& Kirkpatrick et al. (\cite{kirk99})	\\
2M0326		& \object{2MASSW J0326137$+$295015}	& 19.17	& L3.5	& 9.1$\pm$0.2	& $< 1.0$			& Kirkpatrick et al. (\cite{kirk99})	\\
2M0345		& \object{2MASSW J0345432$+$254023}	& 16.98	& L0	& $\leq 0.3$	& $< 0.5$			& Kirkpatrick et al. (\cite{kirk99})	\\
2M0913		& \object{2MASSW J0913032$+$184150}    	& 19.07	& L3	& $< 0.8$	& $< 1.0$			& Kirkpatrick et al. (\cite{kirk99})	\\
2M1145		& \object{2MASSW J1145572$+$231730}    	& 18.62	& L1.5	& 4.2$\pm$0.2	& $< 0.4$			& Kirkpatrick et al. (\cite{kirk99})	\\
2M1146		& \object{2MASSW J1146345$+$223053}    	& 17.62	& L3	& $\leq 0.3$	& 5.1$\pm$0.2			& Kirkpatrick et al. (\cite{kirk99})	\\
2M1334		& \object{2MASSW J1334062$+$194034}	& 18.76	& L1.5	& 4.2$\pm$0.2	& $< 1.5$			& Kirkpatrick et al. (\cite{kirk99})	\\
2M1439		& \object{2MASSW J1439284$+$192915}	& 16.02 & L1	& 1.13$\pm$0.05& $< 0.05$			& Reid et al. (\cite{reid00})	\\
SDSS~0539	& \object{SDSSp  J053951.99$-$005902.0}	& 19.04 & L5	& 		&				& Fan et al. (\cite{fan00})		\\
SDSS~1203	& \object{SDSSp  J120358.19$+$001550.3}	& 18.88 & L3	& 		&				& Fan et al. (\cite{fan00})		\\
\object{Calar 3} 	&               		& 18.73	& M9	& 6.5--10.2	& 1.8$\pm$0.4			& Rebolo et al. (\cite{rebolo96})	\\
\object{Roque 11}	& \object{RPL J034712$+$2428.5} 		& 18.75	& M8	& 5.8$\pm$1.0	&				& Zapatero Osorio et al. (\cite{zap99})	\\
\object{Roque 12}	& 				& 18.47	& M7.5	& 19.7$\pm$0.3	& $\leq 1.5$			& Mart\'\i n et al. (\cite{martin98})	\\
\object{Roque 16}	& \object{RPL J034739$+$2436.4}		& 17.79	& M6	& 5.0$\pm$1.0	& 				& Zapatero Osorio et al. (\cite{zap99})	\\
\object{Teide 1} 	& \object{TPL J034718$+$2422.5}		& 18.80	& M8	& 3.5--8.6	& 1.0$\pm$0.2			& Rebolo et al. (\cite{rebolo95})		\\
S~Ori~31	& \object{S Ori J053820.8$-$024613}	& 17.31	& (M6.5)	&	&				& B\'ejar et al. (\cite{bejar99})		\\
S~Ori~33	& \object{S Ori J053657.9$-$023522}	& 17.38	& (M6.5)	&	&				& B\'ejar et al. (\cite{bejar99})		\\
S~Ori~34	& \object{S Ori J053707.1$-$023246}	& 17.46	& (M6)	& $\leq 5.0$&				& B\'ejar et al. (\cite{bejar99})		\\
S~Ori~44	& \object{S Ori J053807.0$-$024321}	& 19.39	& M6.5	& 60.0$\pm$1.0	&				& B\'ejar et al. (\cite{bejar99})		\\
S~Ori~45	& \object{S Ori J053825.5$-$024836}	& 19.59	& M8.5	&		&				& B\'ejar et al. (\cite{bejar99})		\\
S~Ori~46	& \object{S Ori J053651.7$-$023254}	& 19.82	& (M8.5)	&	&				& B\'ejar et al. (\cite{bejar99})		\\
\hline
\end{tabular}
\end{table*}

Our sample consists of both L dwarfs and late M dwarfs. The targets
were chosen on the basis of being (a) observable for a large fraction
of the night in one observing run, and (b) sufficiently bright that a
good SNR (signal-to-noise ratio) could be achieved in a short
integration time (see section~\ref{observations}).  Within these
selection constraints, we then attempted to observe objects with a
range of spectral types.  Details of the 21 observed objects are given
in Table~\ref{targets}. Ten are field L dwarfs. At the time of the
observations, essentially the only available L dwarfs were the 25
listed by Kirkpatrick et al.\ (\cite{kirk99}) (most of which were
discovered by 2MASS, the Two Micron All Sky Survey), plus (for the
most recent observating run only) a handful from the Sloan Digital Sky
Survey (SDSS).  The ages\footnote{Gizis et al.\ (\cite{gizis00})
investigate the dating of L dwarfs based on activity and kinematics.}
of these objects are generally unknown, but are probably of order 1
Gyr.  The other 11 objects in Table~\ref{targets} are cluster
objects. Five are members of the Pleiades (age 120 Myr), of which two
(Teide 1 and Calar 3) are confirmed brown dwarfs, two (Roque 11 and
Roque 12) are probably brown dwarfs, and the last (Roque 16) is very
close to the hydrogen burning limit so its status is uncertain.  The
six remaining objects are candidate members of the $\sigma$ Orionis
cluster, with masses between 0.02 and 0.12\,$M_{\odot}$, part of this
range reflecting the uncertainty in the cluster age of 1--5 Myr.  The
four faintest objects in this last cluster were observed because they
just happened to be in the field of another target.

\section{Data aquisition and reduction}\label{data}

\subsection{Observations}\label{observations}

The data were obtained over three observing periods: January 1999 (AJD
1187.4--1192.8, hereafter 99-01), September 1999 (AJD 1432.8--1436.2,
hereafter 99-09) and February 2000 (AJD 1601.8--1607.2, hereafter
00-02). AJD is an adjusted Julian day\footnote{In paper I, the
observing dates were incorrectly referred to as MJD instead of JD.},
equal to the Julian Day minus 2450000.  In all cases the CAFOS
instrument on the 2.2m telescope at the Calar Alto Observatory (Spain)
was used.  The objects were observed in the I filter because of their
very red optical colours. The 99-01 run used a 1K$\times$1K TEK CCD
with a $9' \times 9'$ field of view; the other two runs used a SITe
2K$\times$2K CCD windowed to a field of view of 9$' \times 11'$ (to
reduce the readout time). In all cases the pixel scale was
0.53$''$/pix.

To ensure a good variability detection efficiency, we decided that the
magnitude error in the target star should be no more than 0.01
magnitudes at each epoch (i.e.\ SNR$>$110).  On the one hand, a long
integration time is required to achieve this high SNR, but on the
other hand a short one is required to ensure we do not ``blur out''
the variable phenomenon we are trying to observe.  A simple
calculation shows that when observing a sinusoidal variation of period
$\tau$ with an integration time of $t$, then a maximum error of
$3.1t/\tau$ (or typical error of $2.2t/\tau$) in units of the
peak-to-peak amplitude is introduced (provided $\sin (\pi t / \tau)
\simeq \pi t / \tau$).  Tolerating a maximum blurring error of 0.2,
and assuming that no period of interest is below 1--2 hours, we arrive
at a maximum\footnote{Such an integration time does not automatically
preclude detection of shorter periods, because multiple observations are
made at different parts of any sinusoidal curve, but the sensitivity
to very short periods may be reduced.} integration time of around 4--8
minutes. A constant integration time of five minutes was used during
99-01, later increased to eight minutes for the subsequent two
runs. The only exception was the brighter target 2M1439, for which an
integration time of 80s was used to avoid saturation. The eight minute
integration time then set the faintest magnitude limit of the targets
at around $I=19.0$.  Within each night, objects were observed in a
repetitive cycle, although not all objects were observed every
night. During 00-02, two images of the same target were often taken in
each cycle.

\subsection{CCD processing}\label{ccdproc}

The data from 99-01 were presented in paper I, but have been re-reduced
for the present paper in exactly the same manner as the other two
runs. The reduction procedure is now described.

\subsubsection{Basic reduction}

A one-dimensional bias was subtracted from each frame using the
overscan region in each frame.  A small residual two-dimensional bias
pattern remained, and this was removed by subtracting a low-order fit
to a median-combination of many zero-length dark exposures.  The
variable sensitivity across the detector was corrected using
illumination-corrected dome flats, in the following way. Several dome
flats taken through the same optical path (i.e. no telescope or lamp
movemement) were averaged with outlier clipping.  While this is
sufficient to remove the small scale pixel-to-pixel variations, it
will not correct the large scale variations, on account of the
different illuminations from the dome wall and the night sky. Thus the
global illumination of this combined dome flat was removed (by
dividing it by its own low-order fit) and replaced with the global
illumination of the sky. This global sky illumination was obtained by
making a low-order fit to a median-combination of a {\em large number}
(typically 40--50) of night sky images of {\em different
fields}. (These frames were selected from the science frames plus a
number of images of dark patches of sky using the same integration
time.)  These emphasised points are necessary to ensure that bright
stars are removed and do not distort the fit.  The resulting corrected
dome flat is normalised to have unit mean, and each science frame
frame divided by it. This procedure was done separately for each night
of each run.

\subsubsection{Fringe removal}

Most frames showed interference fringes caused by narrow line emission
from the Earth's atmosphere interfering in the non-uniformly-thick
layer on the CCD. The flux amplitude was typically 2\% of the sky
level (a few times the sky noise), and the spatial scale of order a
hundred pixels, so it was essential that these fringes be removed.
Within a given observing run the fringe pattern was found to be
stable, i.e.  independent of time or telescope pointing.\footnote{The
telescope re-pointing accuracy was often good only to tens of pixels,
meaning that in a sequence of frames the image of a given field moved
relative to the fringe pattern.}  It is important to realise that
fringing is an {\em additive} phenomenon. Thus the fringes must be
subtracted from the science frames; they must not be divided, e.g.\
using the flat field, as they have not modulated the star
light. Similarly, the flat field itself must not have fringes, and
it was for this reason that twilight flats could not generally be
used.  The fringes were removed by constructing a ``fringe
correction'' frame, which is a median combination of a set of
flat-fielded night sky images (the same set as used for creating the
sky illumination).  Taking the median at each pixel is necessary to
remove the stars, but this works only if all the frames have a common
flux zero point with respect to these stars, i.e. are sky-subtracted:
Due to varying airmass or the presence of thin cloud and the Moon, the
sky level often differed. Thus before combination, a low-order fit to
each frame was subtracted. The resulting fringe correction frame
showed only the fringes, but was fairly noisy. This was improved with
a spatial smoothing (a boxcar filter of size three pixels). To first
order, the scale of the fringes in a frame is proportional to the
exposure time, so subtracting the smoothed fringe frame usually
removed the fringes. However, in some cases this over- or
under-subtracted the fringes, presumably because the strength of the
fringe pattern also depends on both the airmass and degree of (thin)
cloud cover. In these cases a factor of the smoothed fringe frame was
subtracted, the factor (in the range 0.3 to 2.8) determined manually.

\subsubsection{Error sources}\label{errorsources}

As will be seen in section~\ref{photometry}, an accurate knowledge of
the photometric errors (or at least, not an underestimate) is required
for the detection of variability. For our brighter objects, the
quality of the flat field and the fringe removal set a limit to the
photometric precision.  Through various tests we determined that these
contribute random errors in the photometry of no more than
0.5\%. Other effects which are significantly less could be ignored
(see paper I). Non-linearity in the response of the detector was checked
and could be ignored for flux levels of interest.  A spatial
non-linearity due to the shutter was avoided by using sufficiently
long integration times ($>20$\,s) in all frames. The CAFOS instrument
suffers from geometric distortion, specifically a change in the pixel
scale with distance from the optical axis. As the different images of
a target field were not always identically positioned with respect to
this axis, this potentially introduces errors into relative
photometry. While it can be corrected for, it was found that it
contributed an error in the relative photometry of no more than 0.1\%.

\subsection{Photometry}\label{photometry}

To reduce sensitivity to temporal variations in the
Earth's atmosphere through which the target must be observed, the flux
of the target is monitored relative to a number of reference stars
in the field. These were chosen according to the following criteria:
\begin{enumerate}
\item{near-Gaussian, near-circular point spread function (PSF)
i.e. not an extended object;}
\item{isolated from other sources;}
\item{present on every frame;}
\item{bright (generally brighter than the target, although in several
cases the target was one of the brighter non-saturated stars).  It was
also ensured that the flux was less than about 75\% of the
saturation of the analogue-to-digital convertor (to avoid
non-linearity).}
\end{enumerate}
Aperture photometry was performed on the target and reference stars in
each frame. The choice of aperture size was discussed in paper I.
Although the photometry (and light curve analysis) was done in a range
of aperture sizes, results are presented using an aperture radius of
3.5 pixels, which maximises the SNR while reducing all systematic
errors below the 0.5\% level discussed in section~\ref{errorsources}.

A differential light curve for the target was calculated as follows.
Let $F_{i}$ be the flux (in collected electrons) in the $i^{th}$
reference star of $N$ in a frame.  The reference flux in that frame is
defined as
\begin{equation}
F_r = \frac{1}{N}\sum_i^N F_i
\label{refflux}
\end{equation}
and the reference magnitude is
\begin{equation}
m_r = -2.5\log_{10}F_r
\end{equation}
The relative magnitude of the target is then defined as
\begin{equation}
m_d = m_s - m_r = 2.5\log_{10}(\frac{F_r}{F_s})
\end{equation}
where $F_s$ and $m_s$ are the flux and magnitude of the target
respectively. We
chose to form $m_r$ by averaging fluxes rather than magnitudes, as
this gives more weight to the brighter, higher SNR objects: Simply
averaging magnitudes gives almost as much weight to the faintest
reference stars ($I$\,$\sim$\,19) as to the brightest
($I$\,$\sim$\,16). More ``sophisticated'' weighting schemes did not
create a more precise reference light curve. If $m_d(k)$ is the
relative magnitude in frame $k$, then the light curve is
$m_d(1),m_d(2),\ldots,m_d(k),\ldots,m_d(K)$, from which the mean is
subtracted so that $\sum_k m_d(k) = 0$.

We assume that changes in atmospheric transparency equally affect all
stars\footnote{This is reasonable if the integration time is long
enough for any thin clouds to move across the whole
field of view. See also section~\ref{genresults}.}, so if $m_r$ is defined
using non-variable reference stars (see section~\ref{chisqtest}),
changes in $m_d$ are either due to noise or to intrinsic changes in
the luminosity of the target.  To distinguish between these it is
important to know the errors in $m_d$ as accurately as possible. It
can be shown that the expected error, $\delta m_d$, in $m_d$ is given
by
\begin{equation}
(\delta m_d)^2 = (\delta m_s)^2 + (\frac{1}{NF_r})^2 \sum_i^N F_i^2 (\delta m_i)^2
\end{equation}
where $\delta m_s$ and $\delta m_i$ are the magnitude errors in the target
and $i^{th}$ reference star respectively. The error contributions were
discussed in paper I, but include: noise in the object (assumed to be
Poissonian); noise in the sky (measured from the standard deviation in
the sky aperture); uncertainty in the subtracted sky level; a
contribution of 0.5\% from the ``informal'' errors. This last error
(dominated by imperfect flat fielding and fringe removal) may be a
conservatively large estimate, but ensures increased caution in
claiming to have detected variability.

\section{Time series analysis}\label{timeseries}

\subsection{$\chi^2$ test}\label{chisqtest}

A general test of variability can be made using a $\chi^2$ test, in
which we evaluate the probability that the deviations in the light
curve are consistent with the photometric errors.  The null
hypothesis for this test is that there is no variability\footnote{As
the mean has been subtracted from the light curve, the degrees of
freedom for the test is $K-1$, where $K$ is the number of points
in the light curve.}. We evaluate
\begin{equation}
\chi^2 = \sum_k^K \left( \frac{m_d(k)}{\delta m_d(k)} \right)^2
\end{equation}
and determine the probability, $p$, (from tables) that the null hypothesis is
true. A large $\chi^2$ indicates greater deviation compared to the
errors, and thus a smaller $p$.  We claim evidence for variability if
$p<0.01$.  This test was first used to remove any variable reference
stars, by forming the light curve of each reference star relative to
all the others (see paper I). The test was then applied to the target star
using the resulting non-variable reference stars.

\subsection{Power spectrum estimation with {\sc CLEAN}}

Evidence for {\em periodic} variability was then searched for using
the {\em power spectrum} or periodogram.  In particular, a dominant
periodicity may be present at the rotation period due to rotational
modulation of the light curve by surface inhomogeneities. For a
continuous light curve $g(t)$, the power at frequency $\nu$
is $|G(\nu)|^2$, where
\begin{equation}
G(\nu) = FT[g] = \int_{- \infty}^{+ \infty} g(t) e^{-2\pi i \nu t} \ dt
\label{powerdef}
\end{equation}
and $FT[g]$ denotes the Fourier transform of $g(t)$.  In paper I, this
was estimated using the Lomb--Scargle periodogram, partly on the basis
of the existence of a convenient significance test. However, one of
the drawbacks of this method is that it makes no attempt to remove the
{\em spectral window function} from the data. Suppose we observe
$g(t)$ at certain epochs $t_1,t_2,\ldots,t_n$, specified by the
(discrete) sampling function $s(t)$. The observed data are then given
by the (discrete) function $d(t) = g(t)s(t)$ (which in our case is
just the set $m_d(k)$). The power spectrum we observe is then $|D(\nu)|^2$,
given by
\begin{equation}
D(\nu) = {\rm FT}[d(t)] = G(\nu) \otimes W(\nu)
\label{conv}
\end{equation}
where $W(\nu)={\rm FT}[s(t)]$ is the spectral window function and
$\otimes$ is the convolution operator.  Hence we do not observe the
power spectrum of the process we are observing, but rather the power
spectrum of its convolution with the window function. This can have
serious consequences, as peaks in the power spectrum may be due to the
way in which the data were sampled, and not intrinsic to the observed
process itself (see, for example, Deeming~\cite{deeming75}, Roberts et
al.~\cite{roberts87}). For equal spaced sampling, this manifests
itself as aliasing. For other samplings, $W(\nu)$ can have
considerable ``power'' at a range of frequencies, and for low SNR data
can lead to periodicities in $G(\nu)$ being completely obscured in
$D(\nu)$. One approach to mitigating the effects of the window
function is a deconvolution of $D(\nu)$. Although a direct
deconvolution is not possible, Roberts et al.~(\cite{roberts87})
modified the {\sc CLEAN} algorithm (used to reconstruct
two-dimensional images from interferometric data) to iteratively
remove the spectral window function from the raw, or {\em dirty},
power spectrum.  This works by identifying peaks in the power spectrum
and subtracting the power due to the convolution of $W(\nu)$
associated with them. The resulting {\em cleaned} power spectrum,
$P(\nu)$, generally consists of peaks at a number of distinct
frequencies, plus a residual spectrum consisting of the noise and any
spectral features not well represented by the cleaned frequency
components.

We have used a {\sc CLEAN} algorithm written by Harry Lehto (2000,
private communication).  The cleaned power spectrum is a frequency
domain representation of the light curve, $g(t)$, using sinusoids of
amplitude $A$ (not peak-to-peak), frequency $\nu$ and phase $\phi$,
determined by {\sc CLEAN}.  The power, $P$, at a certain frequency is
related to the amplitude by $A=2\sqrt{P}$ in the noiseless case.  For
evenly spaced data, the noise in the power spectrum (in units of
mag$^2$) is approximately $\overline{\delta m_d}^2/K$, where
$\overline{\delta m_d}$ is the average photometric error and $K$ the
number of points in the light curve.  For a light curve with large
occasional gaps, this result needs to be multiplied by a factor $1 -
(t_{\rm gaps}/t_{\rm max})$, where $t_{\rm max}$ is the total duration
of the light curve and $t_{\rm gaps}$ is the sum of the duration of
the gaps. Peaks which are not more than several times
this noise level should not be considered significant. Note that it is
possible to detect a sinusoid of amplitude less than the photometric
errors, because the noise is spread over many frequencies in the power
spectrum.

We can reasonably search for sinusoidal periods up to the longest time
span of the observations, \tmax, although if the coverage is very
non-uniform then the sensitivity to the longer periods will be
reduced.  There is, in principle, information in the light curve on
periods down to the smallest time separation between epochs.  However,
as the typical spacing between epochs is often more than this, the
sensitivity at these very short periods is similarly reduced.  In
section~\ref{results} we search for periodicities between 0.4 hours
and 125 hours (frequencies between 2.5/hr and 0.008/hr).  The
uncertainty in a period is set by the finite resolution of the power
spectrum. This is determined by the duration of the observations
(\tmax), which makes it impossible to distinguish between two closely
separated frequencies, giving rise to an error in a period $\tau$ of
$\tau^2/(2 t_{\rm max})$ (Roberts et al.~\cite{roberts87}).
However, at very short periods, we place a lower limit on the temporal
resolution due to the finite integration time.

It is useful to plot the light curve phased to any significant periods
to ensure that similar variations are not seen in the reference
stars. However, as will be seen in section~\ref{simulations}, the
absence of {\em sinusoidal} variation in the target star does not mean
that this is not a true periodicity.  A more useful (but not
foolproof) check of whether a periodicity is intrinsic to the target
is to calculate the cleaned power spectrum of the reference stars.
Strong peaks present in both this and the power spectrum of the target
may not be intrinsic to the target.  Note that this cannot be done
reliably with the dirty power spectrum: we see from
equation~\ref{conv} that any ``false'' peaks in the dirty power
spectrum, $D(\nu)$, are due to the convolution of the spectral window
function, $W(\nu)$, with the true power spectrum, $G(\nu)$. While both
target and reference stars have the same $W(\nu)$, they have different
$G(\nu)$, so false peaks which appear in the dirty spectrum of the
target will not necessarily be in the dirty spectrum of the reference
stars. They should, however, both be absent in the cleaned spectra.

\subsection{Other methods}

We briefly investigated the phase dispersion minimization method of
Cincotta et al.~(\cite{cincotta95}) for detection of periodic
variability. This method phases the light curve to a range of periods,
and measures the appropriateness of the period using the Shannon
information entropy in the amplitude--phase space. Periodicities in
the data give rise to minima of the information entropy.  It was found
that the most significant minima were due to the sampling, with
dominant minima at 24 hours and rational multiples thereof. The
numerous other minima were weak and obscured by noise. It seems that
this method may not be suitable for time series with the relatively
few number of points used here (Cincotta, private communication).
This method has not, therefore, be pursued in any detail in this
paper.

\section{Results}\label{results}

\subsection{General results}\label{genresults}

\begin{table*}
\caption[]{Variability detections. $t_{\rm max}$ is the maximum time
span of observations: the minimum span was between 10 and 20 minutes.  The
amplitude of the observed variability is measured by the average (over
all points in the light curve) of the absolute relative magnitudes,
$\overline{|m_d|}$, and the RMS (root-mean-square) of
the relative magnitudes, $\sigma_m$.  $\overline{\delta m_d}$ is the
average photometric error in the light curve (also in
magnitudes). $1-p$ is the probability that the variability is not
compatible with the photometric errors. ``Obs.\ run'' refers to which
of the three observing runs the data come from, in YYMM date format.
\label{detections}
}
\begin{tabular}{llrrrrrrrr}
\hline
\vspace*{0.2ex}
target		& SpT	& \tmax	& $\overline{|m_d|}$	& $\sigma_m$	& $\overline{\delta m_d}$	& $p$	& No.\	& No.\	& Obs.	\\
		&	& hours	&		&		&		&	& frames& refs	& run	\\
\hline
2M0345		& L0	&  53	& 0.012		& 0.017		& 0.011     	&  4e-4	& 27	& 23	& 99-09  \\
2M0913  	& L3    & 125	& 0.042         & 0.055		& 0.039     	&  7e-4	& 36	& 14	& 99-01  \\
2M1145		& L1.5	& 124	& 0.026         & 0.031		& 0.022     	&  1e-3	& 31	& 12	& 99-01  \\
$''$		& $''$	&  76	& 0.015         & 0.020		& 0.012     	&$<$1e-9& 70	& 11	& 00-02  \\
2M1146		& L3	& 124	& 0.012         & 0.015		& 0.011     	&  3e-3	& 29	&  7	& 99-01  \\
2M1334		& L1.5	& 126	& 0.017         & 0.020		& 0.011     	&$<$1e-9& 51	& 12	& 00-02  \\
SDSS~0539	& L5	&  76	& 0.009         & 0.011		& 0.007     	&  3e-5	& 31	& 24	& 00-02  \\
SDSS~1203	& L3	&  52	& 0.007         & 0.009		& 0.007     	&  2e-3	& 51	& 13	& 00-02  \\
Calar 3		& M9	&  29	& 0.026         & 0.035		& 0.027     	&  6e-4	& 42	& 21	& 99-01  \\
S~Ori~31	& (M6.5)	&  50	& 0.010         & 0.012		& 0.007     	&  4e-5	& 21	& 30	& 00-02  \\
S~Ori~33	& (M6.5)	&  51	& 0.008         & 0.010		& 0.007     	&  2e-3	& 21	& 43	& 00-02  \\
S~Ori~45	& M8.5	&  50	& 0.051  	& 0.072  	& 0.032		&  5e-9 &  21	& 30	& 00-02	\\
\hline
\end{tabular}
\end{table*}

\begin{table*}
\caption{Variability non-detections.  The columns are the same as in
Table~\ref{detections} except that here $\overline{|m_d|}$ and
$\sigma_m$ are the upper detection limits on the variability
amplitudes.  The minimum time between observations of a given target
was between 3 minutes (for 2M1439) and 35 minutes (for Roque 12).
\label{nondetections}
}
\begin{tabular}{llrrrrrrrr}
\hline
target		& SpT	& \tmax	& $\overline{|m_d|}$	& $\sigma_m$	& $\overline{\delta m_d}$	& $p$	& No.\	& No.\	& Obs.	\\
		&	& hours	&		&		&		&	& frames& refs	& run	\\
\hline
2M0030		& L2	&  51	& 0.018		& 0.025		& 0.020		& 0.21	& 37	& 27	& 99-09 	\\
2M0326		& L3.5	&  49	& 0.021  	& 0.029		& 0.017		& 0.56	& 19	& 36	& 99-09 	\\
2M1439		& L1	&  97	& 0.007  	& 0.009		& 0.007		& 0.10	& 48	& 13	& 00-02	\\
Roque 11	& M8	& 100	& 0.028  	& 0.043		& 0.027		& 0.46	& 47	& 23	& 99-01	\\
Roque 12 	& M7.5	&  50	& 0.016  	& 0.022		& 0.015		& 0.02	& 17	& 43	& 99-09	\\
Roque 16	& M6	&  29	& 0.010  	& 0.014		& 0.010		& 0.35	& 16	& 34	& 99-09	\\
Teide 1		& M8	& 100	& 0.029  	& 0.041		& 0.030		& 0.10	& 47	& 23	& 99-01  \\
S~Ori~34	& (M6)	&  51	& 0.008  	& 0.010 	& 0.007		& 0.28	& 21	& 43	& 00-02	\\
S~Ori~44	& M6.5	&  51	& 0.030  	& 0.035		& 0.026		& 0.06	& 21	& 30	& 00-02 	\\
S~Ori~46	& (M8.5) 	&  51	& 0.032  	& 0.041 	& 0.030		& 0.03	& 21	& 43	& 00-02	\\
\hline
\end{tabular}
\end{table*}

The results of the application of the $\chi^2$ test to the 21 targets
are shown in Table~\ref{detections} for the detections ($p<0.01$) and
Table~\ref{nondetections} for the non-detections ($p>0.01$) of
variability.  In these tables we use two measures of the variability
amplitude. The first, $\sigma_m$, is simply the RMS (root-mean-square)
value of $m_d(k)$ for the whole light curve (all $k$).  This measure
disproportionately represents large values, so we also quote
$\overline{|m_d|}$, the mean of the absolute values of $m_d(k)$.
Assigning an amplitude in this low SNR regime is non-trivial. For
example, using a slightly different aperture size can give a slightly
different amplitude, because the noise changes. As the same aperture
has been used for all objects (except SDSS~0539) these amplitudes are
at least comparable. In general one needs to determine the amplitude
by solving for a parametrized model, e.g.\ by marginalising over
nuisance parameters in an appropriate Bayesian framework. We are not
prepared to assign such a model at this time, so we simply report
these measures.

For those objects in which we did not detect variability, we have set
upper limits on the amplitude according to what we could have
detected.  This was done by creating a set of synthetic light curves
by multiplying each $m_d(k)$ by $1+a$, for increasing (small) values
of $a$. The amplitude limits were obtained from that synthetic light
curve which gave $p=0.01$ according to the $\chi^2$ test. (The $p$
value quoted in Table~\ref{nondetections} is that from the actual data.)

The reliability of the $\chi^2$ test clearly depends on an accurate
determination of the magnitude errors in the target. We have checked
this by analysing the relative magnitude variations,
$\overline{|m_d|}$, in reference stars of similar brightness as the
target (both before and after rejection of any variables).  We found
that these variations are similar to (and, in particular, no larger
than) the mean error, $\overline{\delta m_d}$, for the respective
target, indicating that we are not underestimating the errors, and
hence not overestimating $\chi^2$ or the significance of a detection.

We point out that the significance of a detection cannot be judged
simply by looking at the ratio of $\sigma_m$ to $\overline{\delta
m_d}$. This ratio is not the ``sigma detection'' level, because the
light curve consists of many points: In the case of a very
large number of epochs, statistically significant fluctuations could
be recognised even if $\sigma_m$ were hardly more than
$\overline{\delta m_d}$. The $\chi^2$ distribution takes this into
account via the degrees of freedom.

Some of the detections/non-detections in the tables are close to the
significance limit, for which a value of $p=0.01$ was chosen as being
reasonably conservative. The choice is, however, somewhat arbitrary,
and we could have chosen 0.05 or 0.001, which would make
some detections into non-detections, or vice versa. We mention this to
emphasise that detections/non-detections close to the limit should be
treated with due uncertainty.

Three of the detections in Table~\ref{detections} (2M0913, 2M1146 and
Calar 3) were non-detections in paper I. These new detections have
amplitudes below the limits placed on them in paper I. Roque 11 and
Teide 1 (non-detections in paper I) remain non-detections, but now at
lower amplitude limits.  The increased sensitivity in the present
paper come about for a number of reasons: improved flat fielding,
including an illumination correction; better fringe removal, including
a lower noise fringe correction image; use of more reference stars; a
slightly smaller photometry aperture to improve the SNR.

In analysing the light curves, it came to our attention that four of
the five targets from 99-09 seem to have a lower average flux on the
last (fourth) night than the average of the preceding three nights
(by 0.01 to 0.03 magnitudes; the fifth target, Roque 12, was not
observed on this night).  This effect is not seen, however, in any of
the reference stars, not even ones of similar brightness to the
targets.  After eliminating other potential problems with the
observing and reduction, one possible cause is that the effective
bandpass was different on this night. As the reference stars are
presumably much bluer on average than the targets, this could change
the magnitude of the targets relative to the reference stars without
changing the magnitudes of the reference stars relative to one
another.  The beginning of the fourth night was lost to cloud and
humidity, and residual cloud cover could have remained for the rest of
the night. However, it appears that a thin cloud layer does not
significantly alter the wavelength dependence of the atmospheric
extinction coefficient over the I band (Driscoll~\cite{driscoll}), so
we cannot provide a satisfactory explanation of this observation at
this time.  Although it is possible that the effect is intrinsic to
all four objects, it is rather suspicious, so we exclude this night
from our analysis and the results presented in the tables.  If this
night were included, 2M0030 and Roque 16 would become detections. No
such correlated behaviour is seen in the targets from the other runs.
Broad band differential photometry can be affected by second-order
colour dependent extinction, even in clear conditions, but due to the
very small airmass gradient across the field-of-view, this
contribution is estimated to be well below the 0.5\% error (see
Young~\cite{young91} for a discussion).

\subsection{Comments on individual objects}

Notes are now given on all the objects with statistically significant
$\chi^2$ detections, along with brief comments at the end of the section on
the non-detections.  The implications of these results will be
discussed in section~\ref{discussion}.\\

\noindent{\em 2M0345}.  The light curve shows no interesting features
and there are no peaks in the cleaned power spectrum above four times
the noise. If the dubious fourth night is included this becomes a very
significant detection ($p <$1e-9). \\

\begin{figure}
\psfig{figure=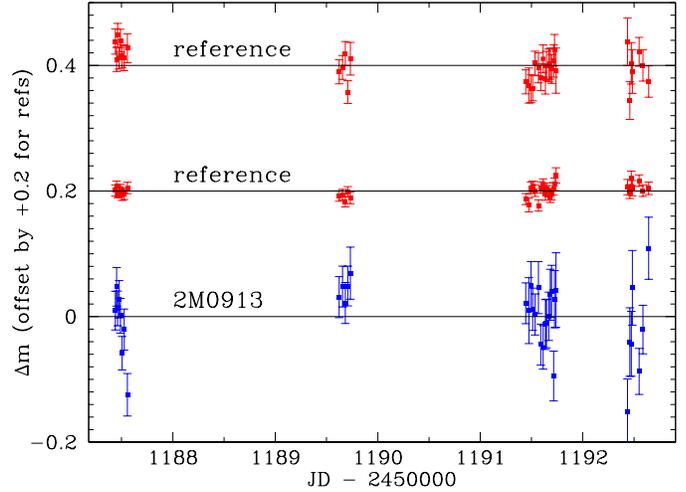,angle=-90,width=0.5\textwidth}
\caption{Light curve for 2M0913 (bottom).  Plotted above this for
comparison are a reference star of similar magnitude (top) and a
bright reference star (middle).  The mean of each light curve is shown
with a solid line. The light curves for the two reference objects are
offset from that for the target star by the amount shown on the
vertical axis.  }
\label{2m0913_lc}
\end{figure}

\noindent{\em 2M0913}. This detection is due primarily to a
significant drop in the flux around AJD 1187.5 (Fig.~\ref{2m0913_lc}),
going down to 0.13 magnitudes below the median for that night, and can
be seen when a range of aperture sizes are used for the photometry.
Although there was some cloud and Moon around this time, no similar
drop is seen in the reference stars, including two of similar
brightness to 2M0913.  Furthermore, two other targets taken at this
time (2M1145 and 2M1146) do not show this behaviour.  There is no
evidence for variability within the other three nights. There are no
strong periodicities in the cleaned power spectrum, the strongest
three being at 3.36, 0.76 and 0.64 ($\pm 0.08$) hours, each at around
only five times the noise level.\\

\begin{figure}
\psfig{figure=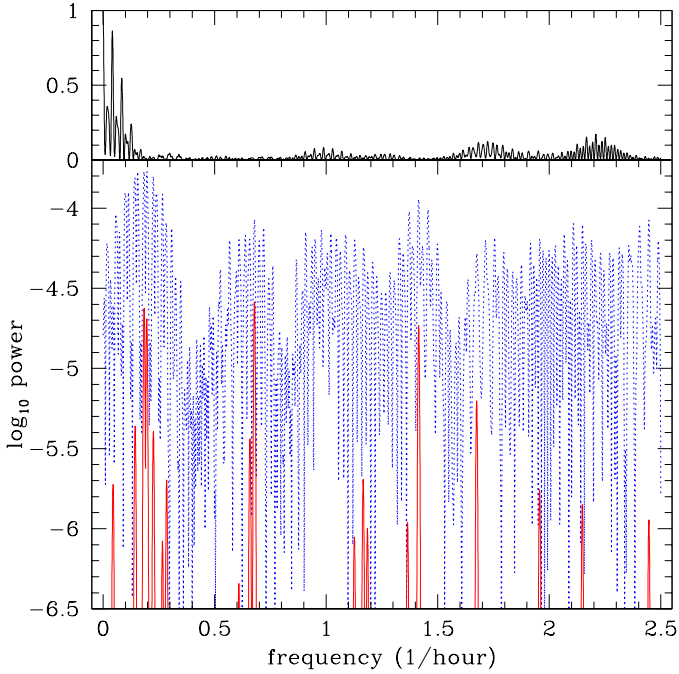,angle=0,width=0.5\textwidth}
\caption{Power spectrum for 2M1145 light curve from the 99-01 run. The
bottom panel shows the dirty spectrum (dotted line) and the cleaned
spectrum (solid line) in units of $\log_{10}(P)$. The noise level is
at about $\log_{10}(P) = -5.6$.  The top panel shows the
spectral window function on a linear vertical scale, normalised to a
peak value of 1.0.}
\label{2m1145_9901_ps}
\end{figure}

\noindent{\em 2M1145}. Evidence for variability in this L dwarf was
presented in paper I, and it was tentatively claimed to be periodic
with a period of 7.1 hours (using the Lomb--Scargle periodogram),
pending confirmation.  The cleaned power spectrum of these same data (old reduction)
gives peaks at $7.1 \pm 0.3$ and $0.71 \pm 0.08$ hours.
The new reduction of these data still gives evidence for variability,
but the cleaned power spectrum shows peaks (all at about eight
times the noise) at $5.4 \pm 0.1$, $5.1 \pm 0.1$, $1.47 \pm 0.08$ and
$0.71 \pm 0.08$ hours (Fig.~\ref{2m1145_9901_ps}).

The improved reduction in the present paper has reduced the average
photometric error from 0.027 to 0.022 magnitudes. (Three additional
frames in the new reduction two nights earlier are also used, which
improves the resolution of the power spectrum.) The light curves from
the two reductions are consistent within their combined errors.  A
small peak is still seen around 7.1 hours in the new reduction, but it
has far less power. In the dirty spectrum, this peak is one of the
strongest, indicating that it has probably been artificially enhanced
by the window function: this demonstrates the necessity of cleaning
the power spectrum.  We see in Fig.~\ref{2m1145_9901_ps} how difficult
it would be to confidently locate the dominant peaks in the dirty
spectrum.  We are confident of the superiority of the new reduction,
so while the variability detection in 2M1145 in paper I still holds,
the tentatively assigned period of 7.1 hours does not.

\begin{figure}
\psfig{figure=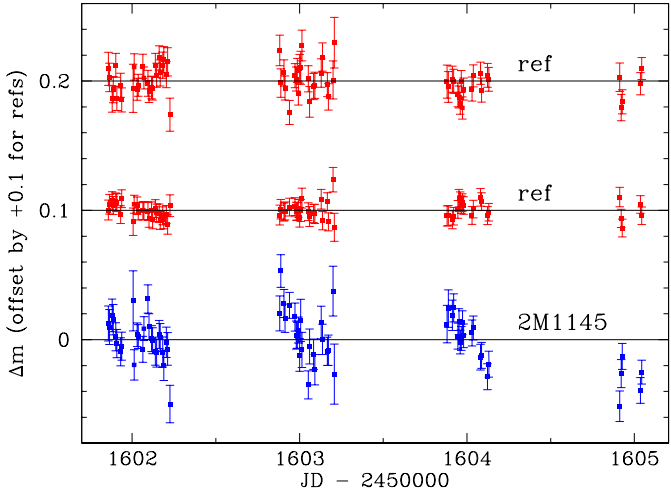,angle=-90,width=0.5\textwidth}
\caption{Light curve for 2M1145 from the 00-02 run (bottom) plus a bright reference object (middle) and one of similar brightness
to the target (top). See caption to Fig.~\ref{2m0913_lc}.}
\label{2m1145_0002_lc}
\end{figure}
\begin{figure}
\psfig{figure=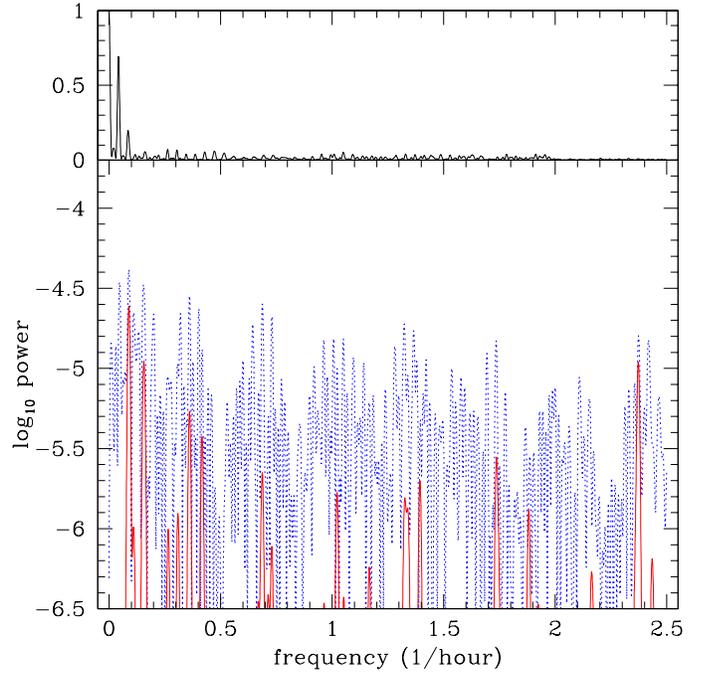,angle=0,width=0.5\textwidth}
\caption{Power spectrum for 2M1145 (from 00-02). The noise level is $\log_{10}(P) = -6.1$. See caption to Fig.~\ref{2m1145_9901_ps}.}
\label{2m1145_0002_ps}
\end{figure}
2M1145 was re-observed at higher SNR and with more epochs across four
nights in the 00-02 run. These data (Fig.~\ref{2m1145_0002_lc}) also
show very strong evidence for variability, and the power spectrum
shows four significant peaks at the following periods (with power in
units of the noise in parentheses): $11.2 \pm 0.8$ (31), $6.4 \pm 0.3$
(14), $2.78 \pm 0.13$ (7), $0.42 \pm 0.13$ (14) hours
(Fig.~\ref{2m1145_0002_ps}).  Note that the first period is four times
the third, so these may not be independent.  There are essentially no
common peaks in this power spectrum and the one from 99-01.  As
mentioned earlier, most epochs in the 00-02 run were taken in pairs
with no time gap between them. This enables us to produce a binned
light curve consisting of 33 points (four single points removed).  The
cleaned power spectrum of this only has a significant periodicity at
$11.3 \pm 0.8$ hours (8 times the noise). There is still a periodicity
at $2.77 \pm 0.30$ hours, but now at only five times the noise
level. It is unlikely that either is the rotation period, as neither
was detected in the 99-01 data (Fig.~\ref{2m1145_9901_ps}). We
can be confident that 2M1145 does not have both {\em stable} (over a
one year timescale) surface features and a rotation period of between
1 and 70 hours. If it did, we would have detected such a rotation
period in both runs (see section~\ref{simulations}).  \\

\begin{figure}
\psfig{figure=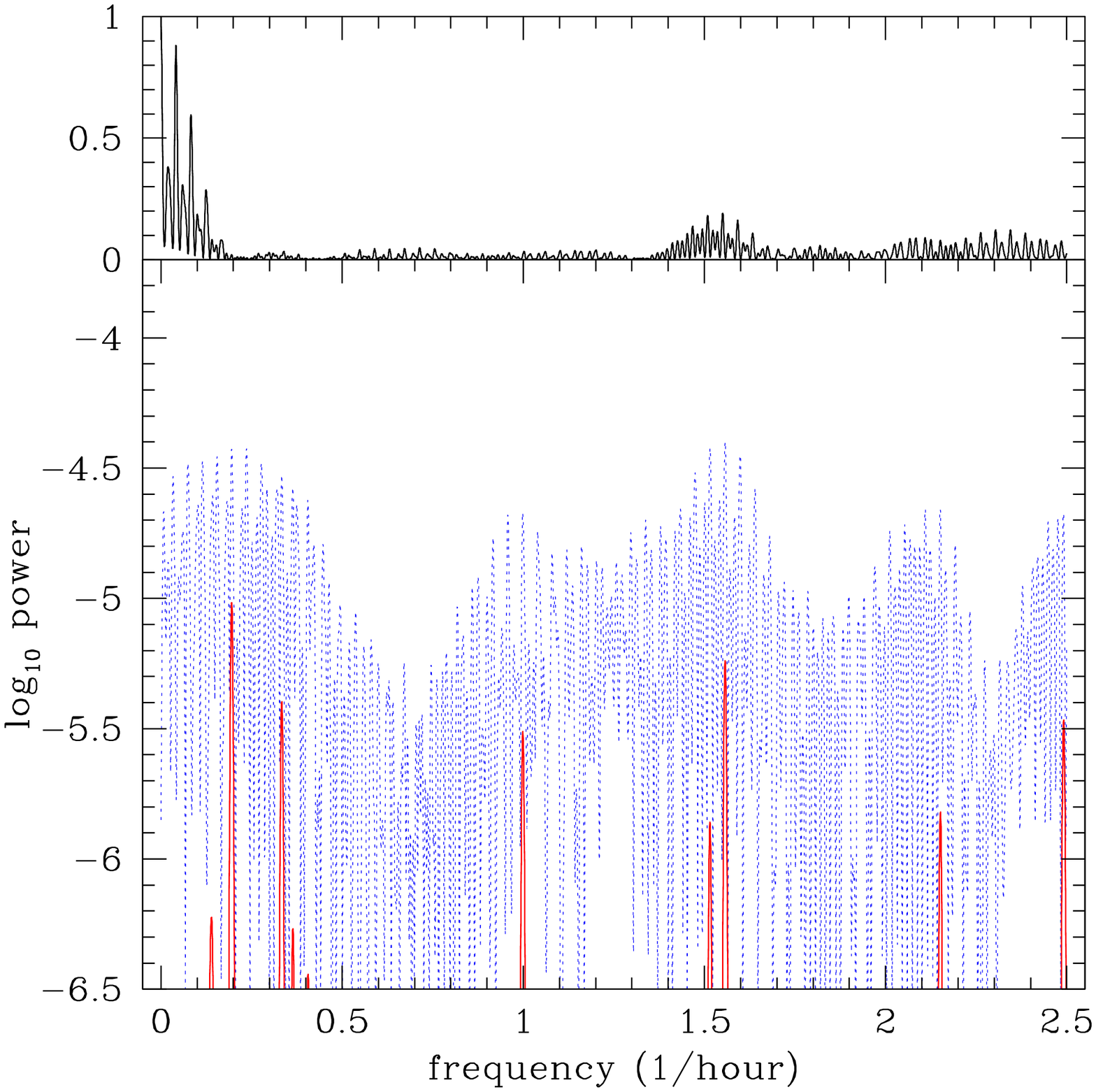,angle=0,width=0.5\textwidth}
\caption{Power spectrum for 2M1146. The noise level is $\log_{10}(P) = -6.1$. See caption to Fig.~\ref{2m1145_9901_ps}.}
\label{2m1146_ps}
\end{figure}
\noindent{\em 2M1146}. This is a marginal detection which was a
marginal non-detection in the original reduction. The power spectrum
shows peaks at the following periods (with power in units of noise):
$5.1 \pm 0.1$ (15), $3.00 \pm 0.08$ (6), $1.00 \pm 0.08$ (5), and
$0.64 \pm 0.08$ (9) hours (Fig.~\ref{2m1146_ps}). The second and third
are in the ratio 3:1, so are probably not independent. The one at
three hours is more convincing based on the phase coverage in the
phased light curve. This is one of only two L dwarfs in our sample
which already has a measured $v \sin i$ of 32.5$\pm$2.5\,\kms\ (Basri
et al.~\cite{basrietal00}). For an object of radius $0.1 R_{\odot}$
(expected for these objects, see Chabrier \& Baraffe
\cite{chabrier00a}), this implies a rotation period of $3.7 \pm 0.3$
hours, or less, due to the unknown inclination, $i$, of the rotation
axis to the line of sight.  In the case of 2M1146, however, there is
another complicating factor, namely that Koerner et al.\
(\cite{koerner99}) have observed it to be a brightness ratio one
binary, with separation 0.3$''$ (7.6 AU)\footnote{The circular orbital
speed about their centre-of-mass is less than 2.5\,\kms\ (assuming
masses of $< 0.1 M_{\odot}$), so does not complicate the $v \sin i$
determination of Basri et al.}.  This was not resolved by our
observations, so our light curve (and power spectrum) is a composite
of the two objects. It is possible, therefore, that two of the three
peaks in the power spectrum are rotation periods for the objects.
Kirkpatrick et al.~(\cite{kirk99}) also found an earlier type star
1$''$ away, which is presumably a background star, and this too could
effect our light curves.  \\

\begin{figure}
\psfig{figure=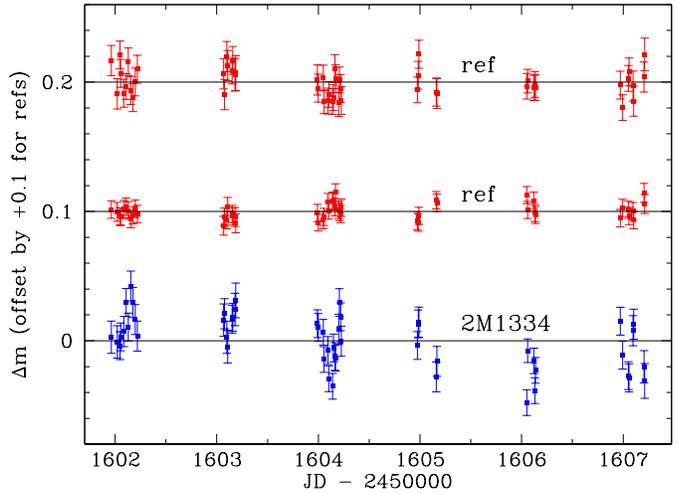,angle=-90,width=0.5\textwidth}
\caption{Light curve for 2M1334 (bottom) plus a bright reference
object (middle) and one of similar brightness to the target (top). See
caption to Fig.~\ref{2m0913_lc}.}
\label{2m1334_lc}
\end{figure}

\begin{figure}
\psfig{figure=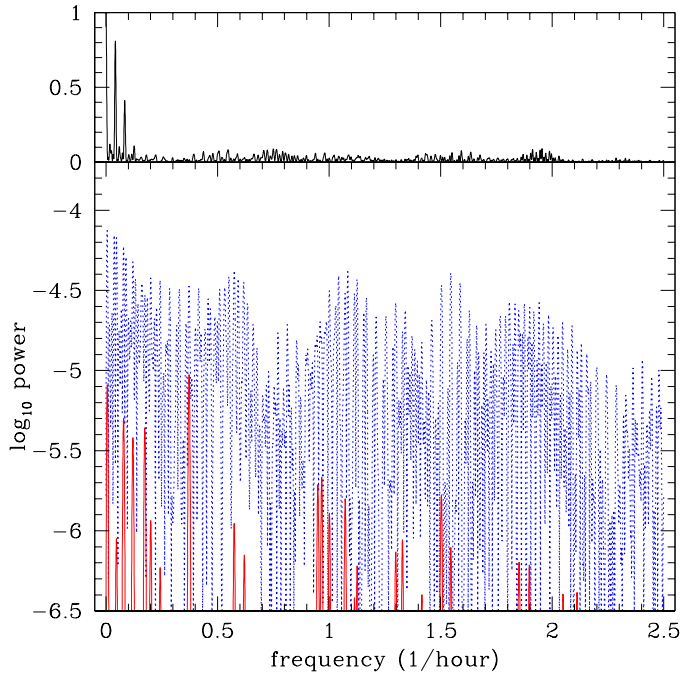,angle=0,width=0.5\textwidth}
\caption{Power spectrum for 2m1334 (all nights). The noise level is $\log_{10}(P) = -6.2$. See caption to Fig.~\ref{2m1145_9901_ps}.}
\label{2m1334_ps}
\end{figure}

\noindent{\em 2M1334}. This is significantly variable, and the light
curve shows clear fluctuations within a number of nights
(Fig.~\ref{2m1334_lc}).  The largest peak in the power spectrum
(Fig.~\ref{2m1334_ps}) is at $2.68 \pm 0.13$ hours at 12 times the
noise. If we look more closely at the raw light curve, the first three
nights would appear to show a periodicity on the scale of a few hours
(the $\chi^2$ value for just these three nights is $p=$\,2e-6).  The
power spectrum of just these three nights shows peaks at $6.3 \pm 0.4$
and $1.01 \pm 0.08$ hours at six and seven times the noise
respectively.\\

\begin{figure}
\psfig{figure=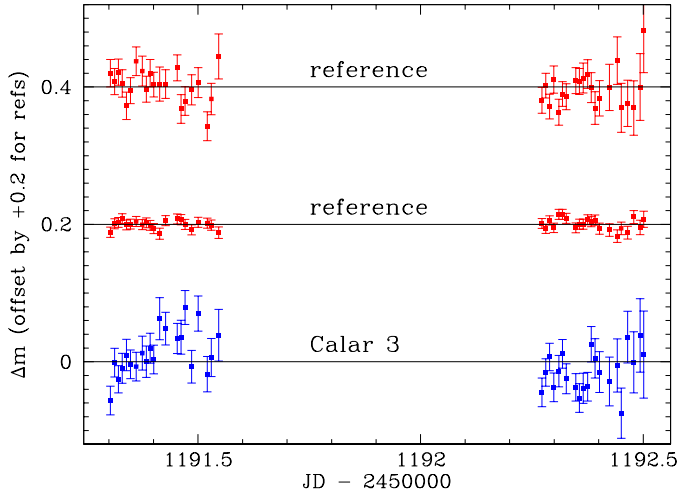,angle=-90,width=0.5\textwidth}
\caption{Light curve for Calar 3 (bottom) plus a bright reference object (middle) and one of similar brightness
to the target (top). See caption to Fig.~\ref{2m0913_lc}.}
\label{calar3_lc}
\end{figure}

\noindent{\em Calar 3}. The light curve (Fig.~\ref{calar3_lc}) does not look
qualitatively different from that of three reference stars of similar
brightness, apart from some ``activity'' around AJD=1191.5.  The two
most significant peaks in the power spectrum (at 14.0 and 8.5 hours)
are less than five times the noise level, so are barely significant.\\

\begin{figure}
\psfig{figure=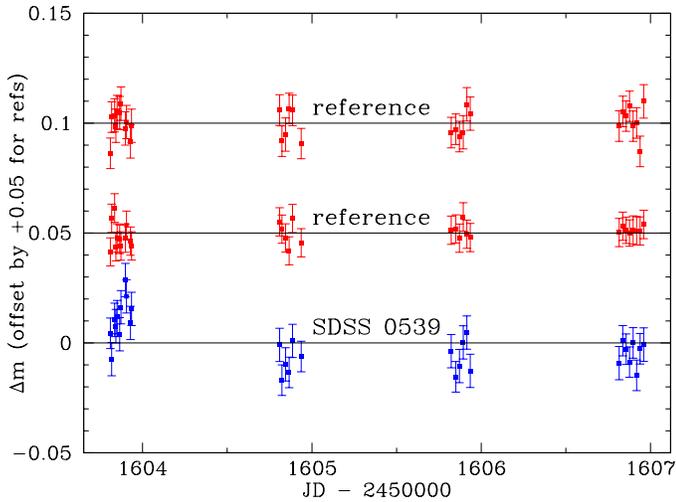,angle=-90,width=0.5\textwidth}
\caption{Light curve for SDSS~0539 (bottom) plus a bright reference object (middle) and one of similar brightness
to the target (top). See caption to Fig.~\ref{2m0913_lc}.}
\label{sdss0539_lc}
\end{figure}

\begin{figure}
\psfig{figure=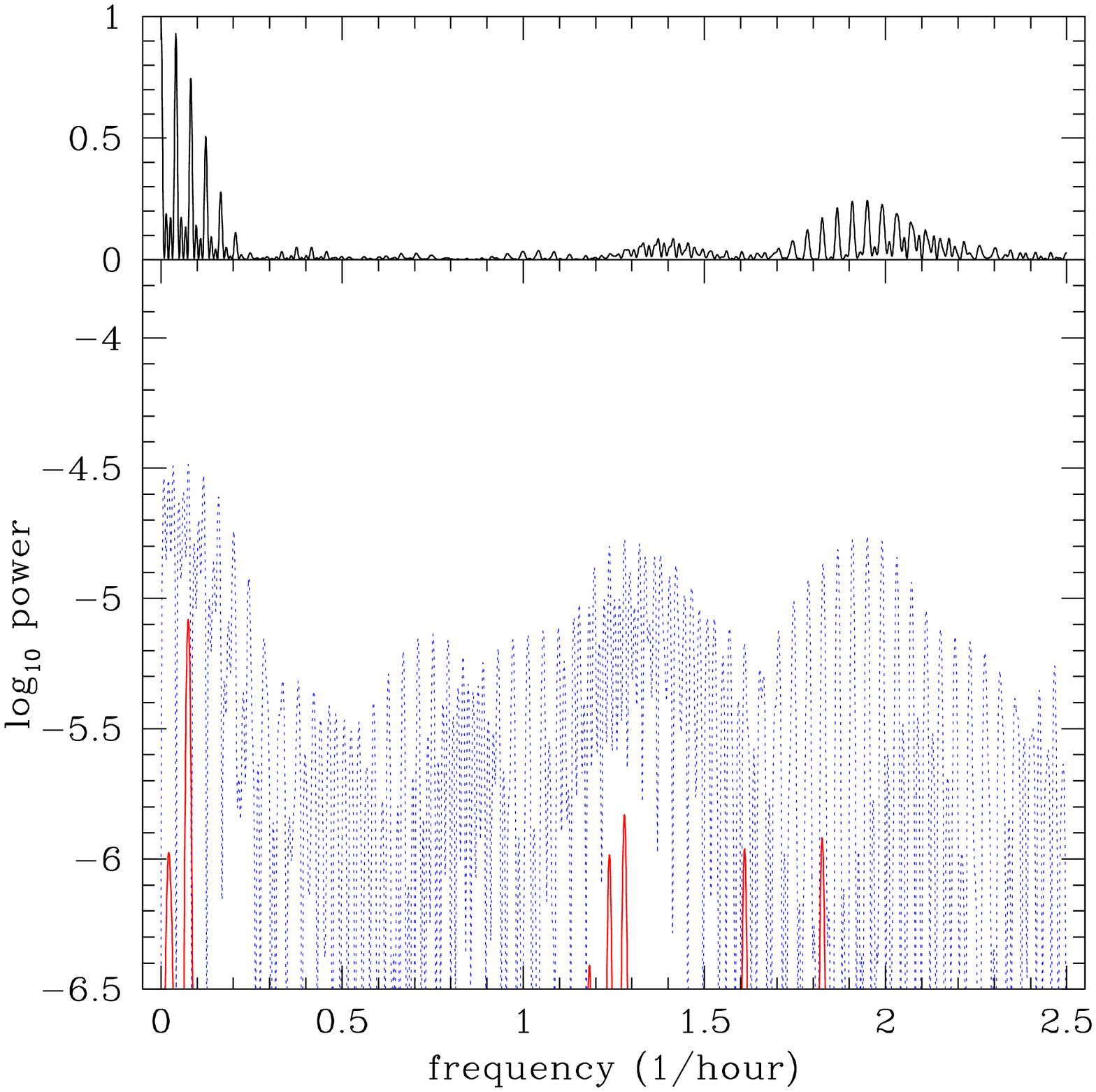,angle=0,width=0.5\textwidth}
\caption{Power spectrum for SDSS~0539. The noise level is $\log_{10}(P) = -6.4$. See caption to Fig.~\ref{2m1145_9901_ps}.}
\label{sdss0539_ps}
\end{figure}

\begin{figure}
\psfig{figure=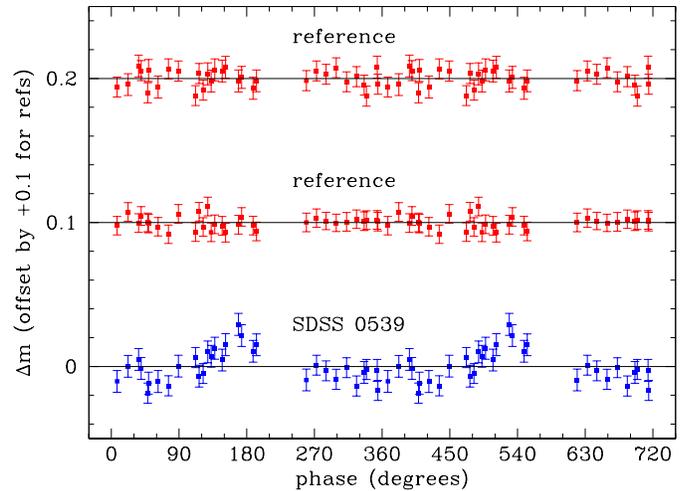,angle=-90,width=0.5\textwidth}
\caption{Light curve (bottom) for SDSS~0539 phased to a period of 13.3
hours. The cycle is shown twice (labelled 0$^{\circ}$--360$^{\circ}$ and
360$^{\circ}$--720$^{\circ}$).  This phased light curve (plus all others in this
paper) has been plotted to have the same phase as a sine wave.  Also
shown are two reference stars phased in the same way.}
\label{sdss0539_ph}
\end{figure}

\noindent{\em SDSS~0539}. The seeing was worse than average for many
of the frames in this field, so a larger photometry aperture of radius
5.0 pixels was used. (Use of a bigger aperture generally decreases the
significance of a detection as it increases the noise, so using a
larger aperture in this case is more conservative.) The significant
$\chi^2$ is partly due to the brighter points around AJD
1604. Otherwise the light curve shows no obvious patterns (see
Fig.~\ref{sdss0539_lc}). The power spectrum shows a significant (20
times noise) peak at $13.3 \pm 1.2$ hours
(Fig.~\ref{sdss0539_ps}). The light curve phased to this period is
shown in Fig.~\ref{sdss0539_ph}. \\

\begin{figure}
\psfig{figure=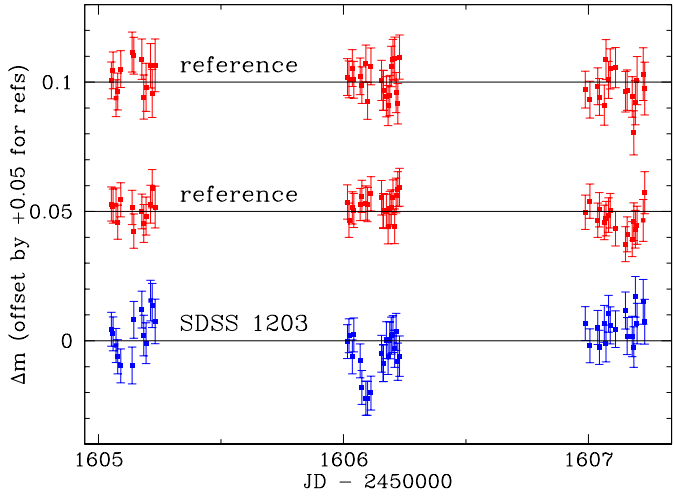,angle=-90,width=0.5\textwidth}
\caption{Light curve for SDSS~1203 (bottom) plus a bright reference object (middle) and one of similar brightness
to the target (top). See caption to Fig.~\ref{2m0913_lc}.}
\label{sdss1203_lc}
\end{figure}

\noindent{\em SDSS~1203}. This variability is primarily due to a drop
in brightness of about 0.02 magnitudes in four consecutive
measurements around AJD=1606.1 (Fig.~\ref{sdss1203_lc}). The drop
lasts between one and two hours.  Particularly interesting
here (as drops in a few consecutive points are often seen) is
that the light curve never drops this low at any other time.  It could
be attributed to an eclipse by a physically associated companion. This
would either have to be very close or of much lower luminosity and
hence mass, possibly a planetary companion.  There are of course other
explanations, such as a short-lived surface feature.  \\

\begin{figure}
\psfig{figure=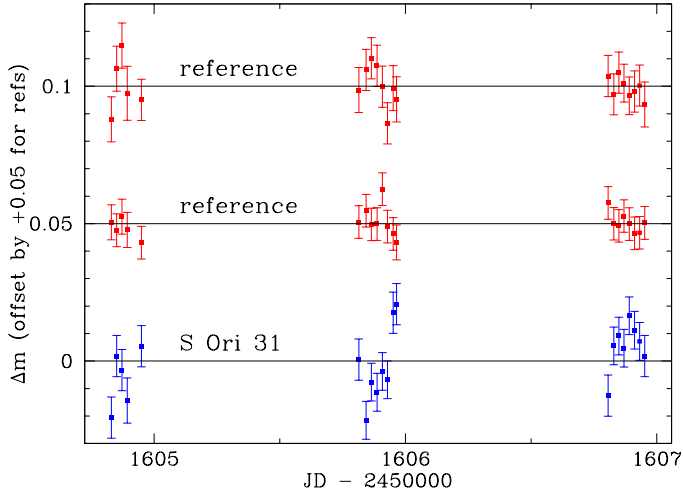,angle=-90,width=0.5\textwidth}
\caption{Light curve for S~Ori~31 (bottom) plus a bright reference object (middle) and one of similar brightness
to the target (top). See caption to Fig.~\ref{2m0913_lc}.}
\label{sori31_lc}
\end{figure}

\begin{figure}
\psfig{figure=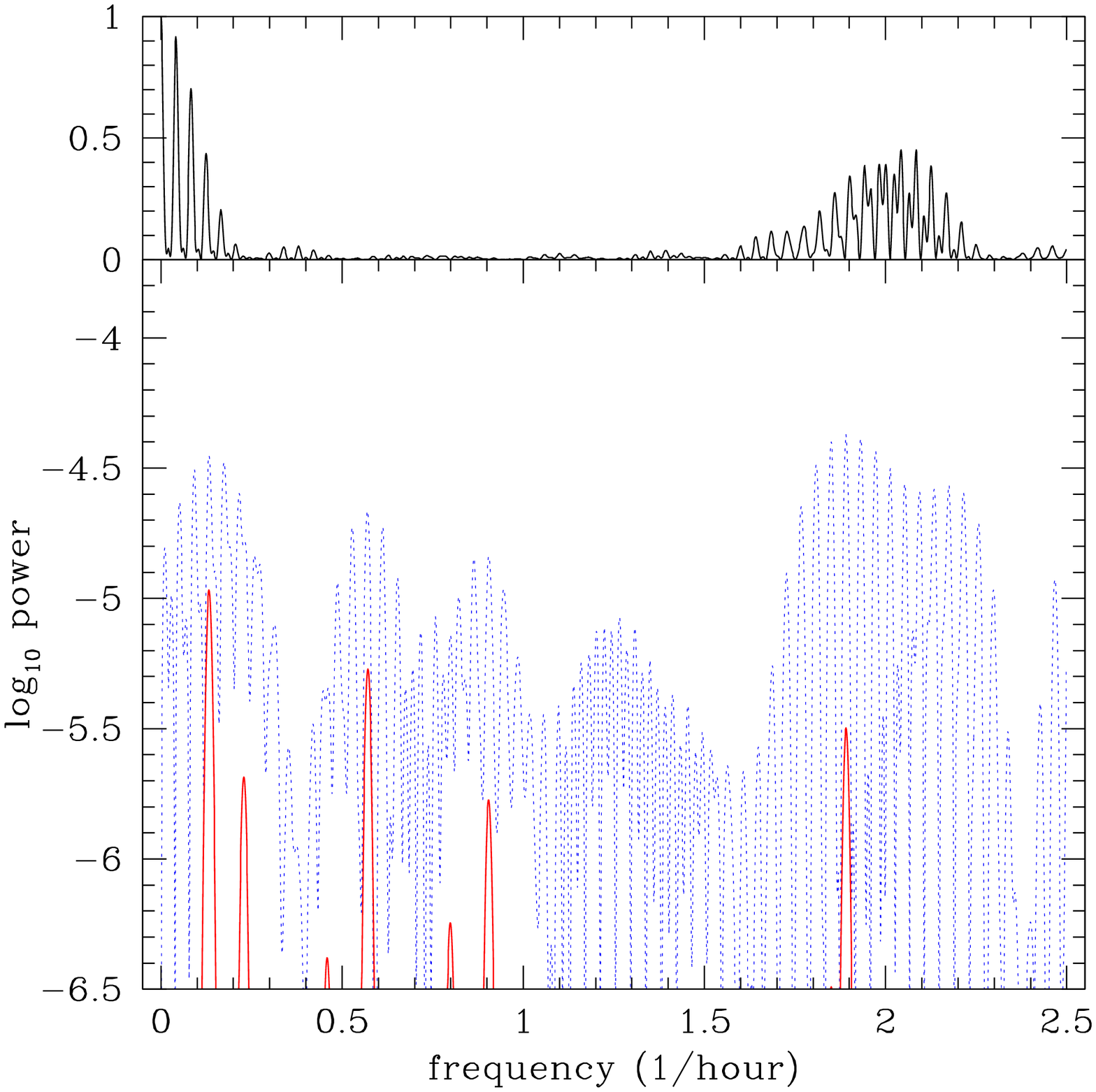,angle=0,width=0.5\textwidth}
\caption{Power spectrum for S~Ori~31. The noise level is $\log_{10}(P) = -6.2$. See caption to Fig.~\ref{2m1145_9901_ps}.}
\label{sori31_ps}
\end{figure}

\begin{figure}
\psfig{figure=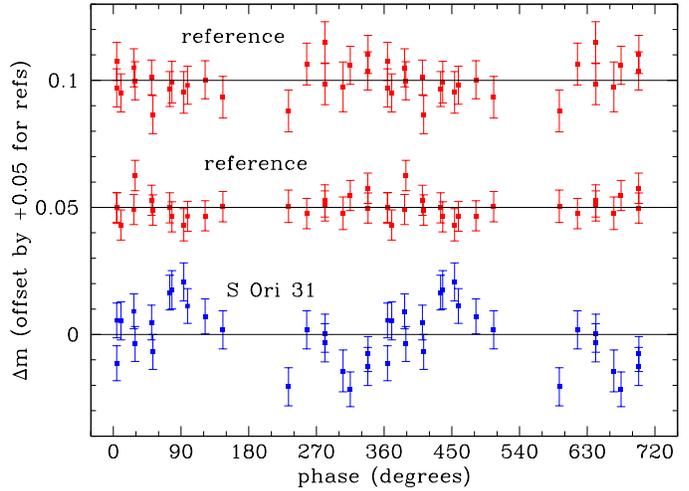,angle=-90,width=0.5\textwidth}
\caption{
Light curve (bottom) for S~Ori~31 phased to a period of 7.5 hours. Also shown are the two
reference stars from Fig.~\ref{sori31_lc}.}
\label{sori31_ph}
\end{figure}

\noindent{\em S~Ori~31}.  The light curve and power spectrum are shown
in Figs~\ref{sori31_lc} and \ref{sori31_ps}.  The latter shows two
significant peaks at $7.5 \pm 0.6$ and $1.75 \pm 0.13$ hours at 18 and
9 times the noise level respectively.  The former period dominates and
shows reasonable evidence for sinusoidal variation
(Fig.~\ref{sori31_ph}), with an amplitude of about 0.01 magnitudes,
and may be the rotation period for this object.\\

\begin{figure}
\psfig{figure=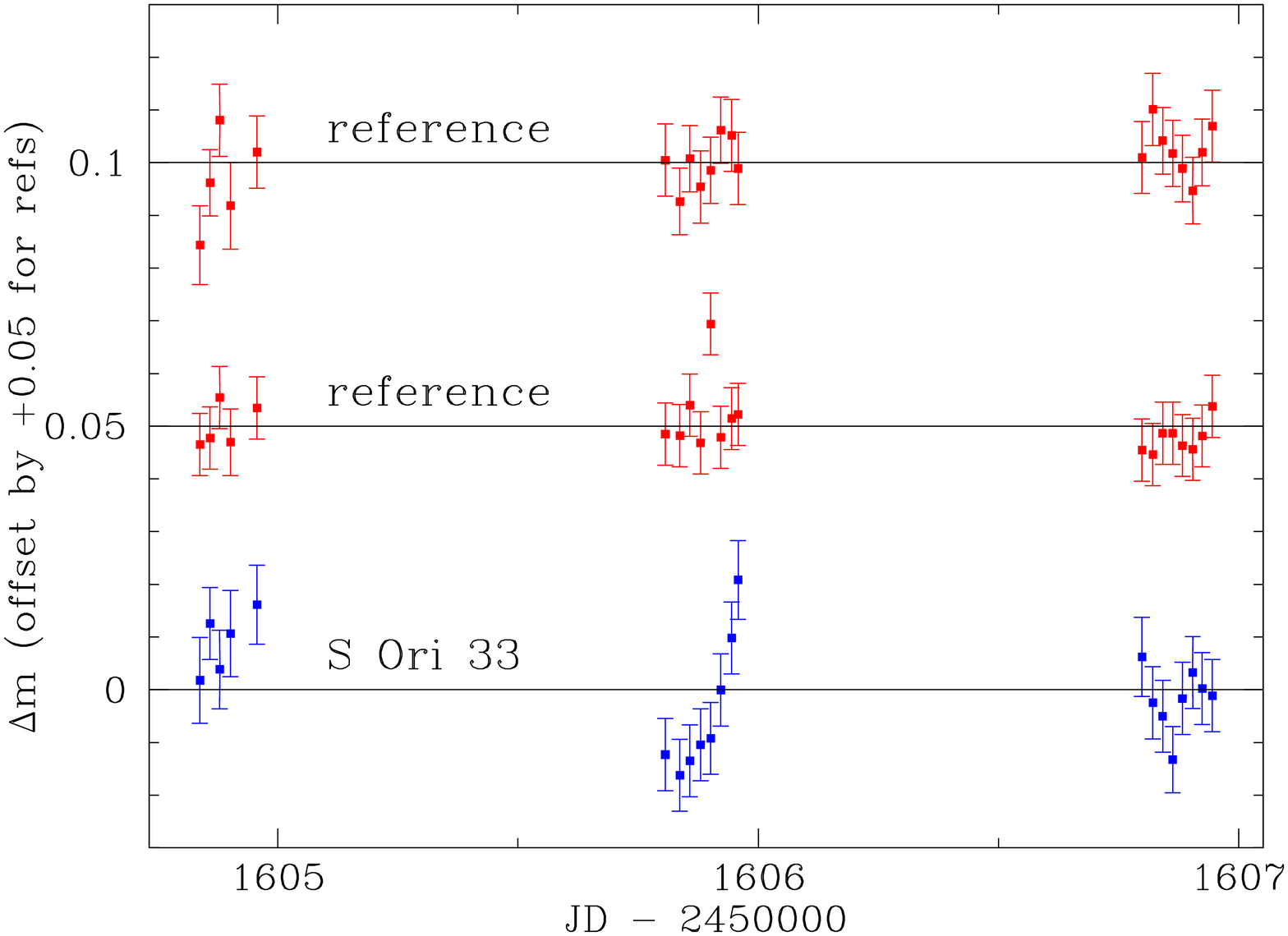,angle=-90,width=0.5\textwidth}
\caption{Light curve for S~Ori~33 (bottom) plus a bright reference object (middle) and one of similar brightness
to the target (top). See caption to Fig.~\ref{2m0913_lc}.}
\label{sori33_lc}
\end{figure}

\begin{figure}
\psfig{figure=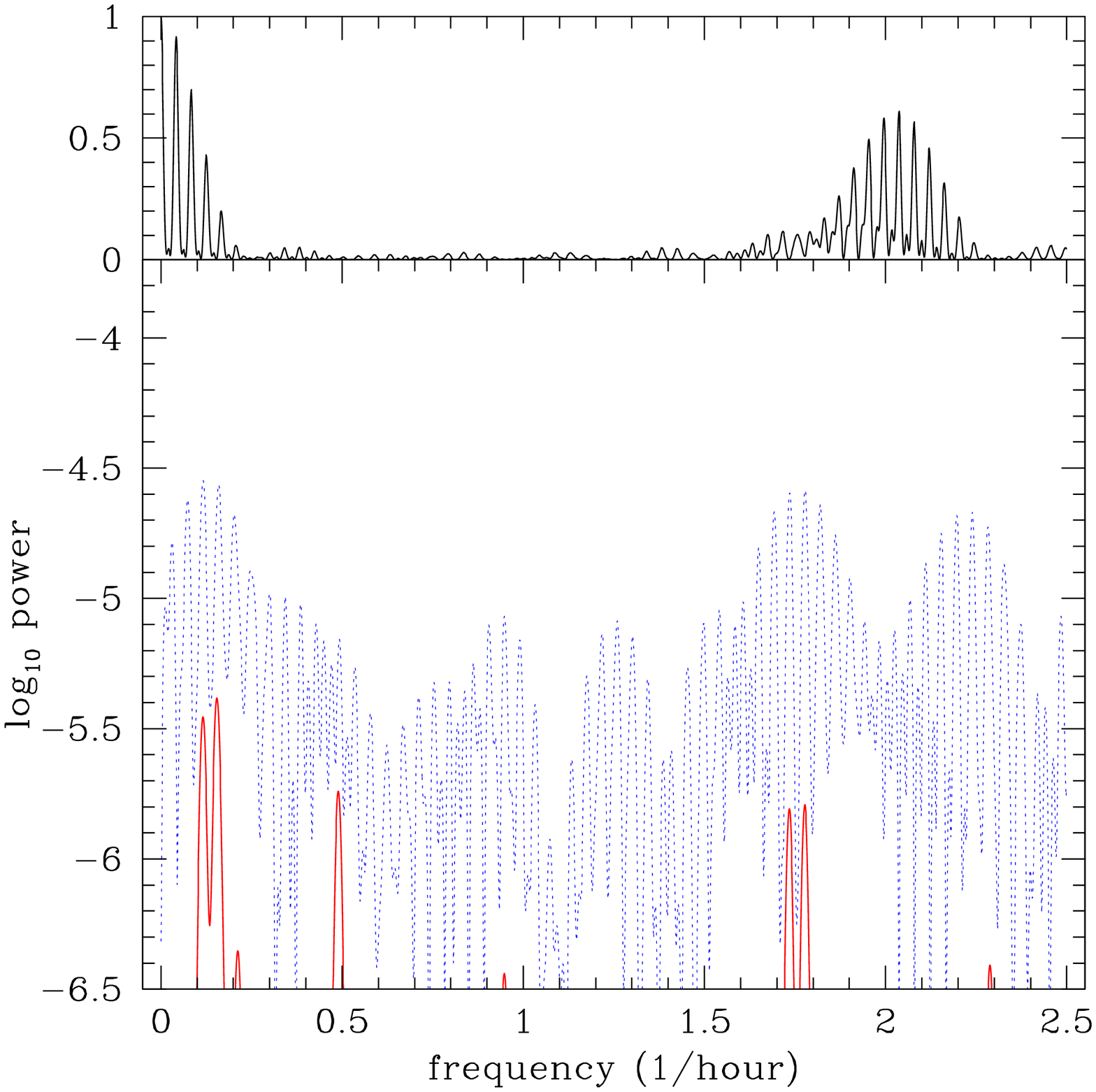,angle=0,width=0.5\textwidth}
\caption{Power spectrum for S~Ori~33. The noise level is $\log_{10}(P) = -6.2$. See caption to Fig.~\ref{2m1145_9901_ps}.}
\label{sori33_ps}
\end{figure}

\begin{figure}
\psfig{figure=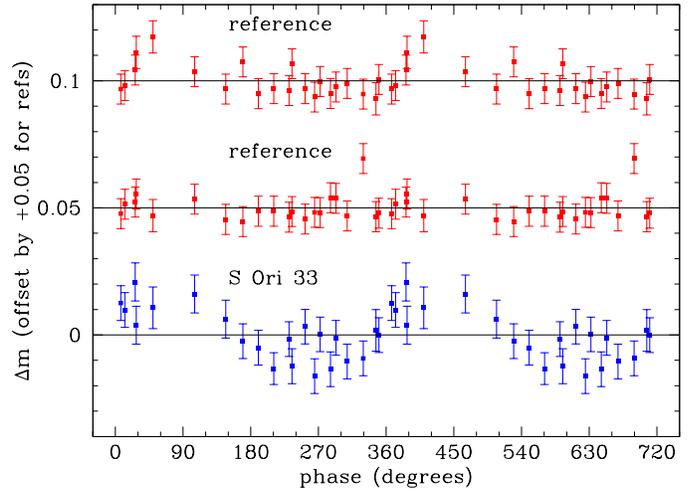,angle=-90,width=0.5\textwidth}
\caption{Light curve (bottom) for S~Ori~33 phased to a period of 8.6 hours. Also shown are the two
reference stars from Fig.~\ref{sori33_lc}.}
\label{sori33_8.6hr_ph}
\end{figure}

\noindent{\em S~Ori~33}. The light curve (Fig.~\ref{sori33_lc}) shows
a rise just before AJD 1606, and the power spectrum
(Fig.~\ref{sori33_ps}) has peaks of 6 to 7 times the noise at $8.6 \pm
0.7$ and $6.5 \pm 0.4$ hours. Although neither is very significant,
the phased light curve at 8.6 hours shows reasonable sinusoidal
variation (Fig.~\ref{sori33_8.6hr_ph}) with an amplitude of around
0.015 magnitudes. This could be the rotation period.  The light curve
phased to 6.5 hours, on the other hand, gives a much poorer fit to a
sine wave.\\

\noindent{\em S~Ori~45}.  The light curve shows three points much
lower than the average around AJD 1604.9.  Indeed, the five points on
this first night of observations span a range of almost 0.25
magnitudes. If these points are excluded there is no evidence for
variability ($p=0.18$).  There is a bright ($\Delta m = 1.7$) star
nearby (5$''$) which may well interfere with this variability
determination. For example, small changes in this star's brightness
could result in large changes in the apparent brightness of S~Ori~45
due to the flux gradient across the sky and photometry apertures of
S~Ori~45. The most significant peak in the power spectrum is at $0.50
\pm 0.13$ hours (at 20 times the noise), which would be extremely fast
if it is the rotation period. Clearly, much more rapid monitoring is
required to determine this. There is a dip in three points around AJD
1606.9, similar to that seen in SDSS~1203, but we are 
hesitant to draw conclusions given the proximity of the bright
star.  S~Ori~44 was observed in the same frame as S~Ori~45, and if we
plot the light curve of one relative to the other, we see that $m_{\rm
sori44} - m_{\rm sori45}$ varies between $+$0.15 and $-0.18$ mag with
a mean of $-0.05$ and a standard error in this mean of $0.01$ mag.
This is interesting, as B\'ejar et al.\ (\cite{bejar99}) give $m_{\rm
sori44} - m_{\rm sori45} = -0.20 \pm 0.08$ mag. While these values are
not inconsistent, the discrepancy could support evidence for
variability in at least one of the objects.\\

\noindent{\em Non-detections}.  2M1439 has been measured by Basri et
al.\ (\cite{basrietal00}) to have a $v \sin i$ of $10 \pm 2.5$\,\kms,
implying a period of less than 12.1 hours for a $0.1 R_{\odot}$
radius.  
S~Ori~44 shows three consecutive points around AJD 1605.9 lower than
the other five points on that night by about 0.09 magnitudes, possibly
indicative of an eclipse, but unlike SDSS~1203 the $\chi^2$ is not
significant (the errors are much larger for S~Ori~44), and on the
following night there are several points at this level. S~Ori~46 has a
bright nearby star, which may affect our attempt to determine
variability. Roque 11 and Teide 1 have also been observed for
variability in the I band by Terndrup et al.\ (\cite{terndrup99}).
They also did not find evidence for variability, with measured values
of $\sigma_m$ (rather than detection limits) of 0.041 and 0.045
magnitudes respectively.

\section{Discussion}\label{discussion}

\subsection{General Comments}

Of the 21 targets observed, 11 show evidence for variability at the
99\% confidence level ($p=0.01$).  Of these, four (2M1145, 2M1334,
SDSS~0539, S~Ori~31) show strong evidence for variability
($p<$\,1e-4). S~Ori~45 is formally a fifth object with strong evidence
for variability, but the presence of a bright close star makes us
hesitant to draw this conclusion. In four cases (2M1146, 2M1334,
SDSS~0539, S~Ori~31) we have detected dominant significant periods in
the range 3--13 hours, which may be rotation periods in all but the
first case.  S~Ori~45 also has a dominant peak, but at 0.5 hours this
would be very rapid if it is a rotation.  The remaining objects do not
show dominant periods, although the two earliest-type variables
(S~Ori~31 and S~Ori~33) show near-sinusoidal light curves at detected
periods. The light curve of one object, SDSS~1203, is essentially
featureless except for a dip which may be due to an eclipse by a
companion, although there is no direct evidence for this.

All of the objects which show variability have RMS amplitudes
($\sigma_m$ in Table~\ref{detections}) between 0.01 and 0.055
magnitudes (ignoring S~Ori~45). The lower limit is set by the
sensitivity of the observations, but no such upper limit is set. Thus
one conclusion from this work is that these objects generally only
have small amplitude variations, most in the range 0.01 to 0.03
magnitudes, on timescales of typically a few to a few tens of
hours. The large fraction of non-detections (50\%), with upper limits
on their RMS amplitudes as low as 0.01 magnitudes, indicates that at
least some ultra cool dwarfs have variability amplitudes less than
0.01 magnitudes.

These detections/non-detections are claimed on the basis of a $\chi^2$
test of the light curves. This requires a careful estimation of the
photometric errors for the target objects: we confirmed that these
were not underestimated via a comparison with the variability level of
stars in the field of similar brightness. Additionally, the use of
many reference stars (from which variables were first eliminated) plus
the conservative assignment of a flat-fielding and fringe-removal
error, gives us good confidence that we have not overestimated the
significance of detections. We highlight that the 99\% confidence
level for the detection of variability is a somewhat arbitrary one:
the division between Tables~\ref{detections} and~\ref{nondetections}
represents a confidence level and not a definitive statement of what
is and what is not variable at a certain amplitude.

\subsection{Rotation and surface features: simulations}\label{simulations}

The power spectrum is a representation of the light curve in the
frequency domain (equation~\ref{powerdef}): $P(\nu)$ is the
contribution of a sinusoid at frequency $\nu$ to the variance in the
light curve $g(t)$.  The goal of this analysis is to see whether the
light curve can be more simply explained in this domain. However, the
presence of a significant peak in the power spectrum does not mean
that this is a long-term periodicity.  After all, {\em any} light
curve, including a random one, can be described in terms of its power
spectrum, so the features in the light curve must appear somewhere in
the power spectrum. The question is whether this description tells us
anything useful about the source.  If we detect just one or two
dominant peaks then it may well be appropriate to describe the light
curve as periodic at the detected period(s).  If, on the other hand,
we detect a large number of peaks, then, given that we have a finite
number of data points, these peaks are less likely to correspond to
true long-term periodicities.

The ideal case of a pure sinusoidal light curve is only produced by
a rotating star if one hemisphere is uniformly darker than the other
and the star is observed along its equatorial plane. A star with a
single small surface feature (``spot'') would show a sinusoidal
pattern (due to a cosine projection effect) only when the spot is on
the observable hemisphere; for up to half of the rotation (depending
on the inclination of the rotation axis) the light curve would be
constant.  A star with two spots would show a yet more complex light
curve, as two, one or no spots are observable at any one time.  While
these light curves will be periodic, they will not be sinusoidal, as
additional sine waves are required to reconstruct the exact shape of
the light curve. Hence the power spectrum of the light curve of a
rotating star will typically consist of several peaks, any number of
which may be significant. Of course, certain spot patterns may give
rise to near-sinusoidal variations, but not necessarily so.  For
example, several of the light curves of Herbst et al.\
(\cite{herbst01}) are periodic but not due to a single sinusoidal
component.

\begin{figure}
\psfig{figure=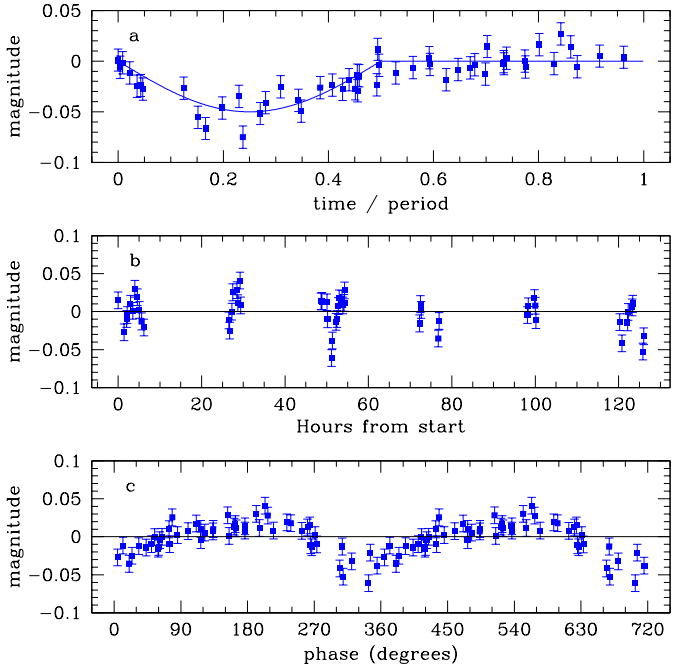,width=0.5\textwidth,angle=0}
\caption{Simulation of the light curve of a spotted rotating star.
{\bf a} The solid line shows the true (noiseless) light curve of
rotating star viewed equatorially with a single dark spot which causes
a dimming of a maximum of 0.05 magnitudes. If the rotation
period is five hours and the star is observed in the same way as
2M1334, i.e. with the same time sampling and Gaussian noise with
standard deviation of 0.011 magnitudes, we obtain the light curve in
{\bf b}, which, when wrapped to the rotation period gives the points
plotted in {\bf a}.  This is significantly variable according to the
$\chi^2$ test ($p<$\,1e-9). A CLEAN power spectral analysis of this
light curve (Fig.~\ref{simul01_ps}) reveals a period of $5.01 \pm
0.10$ hours: the light curve phased to this detected period {\em and
phase} is shown in {\bf c} (cycle shown twice).}
\label{simul01_rot}
\end{figure}

\begin{figure}
\psfig{figure=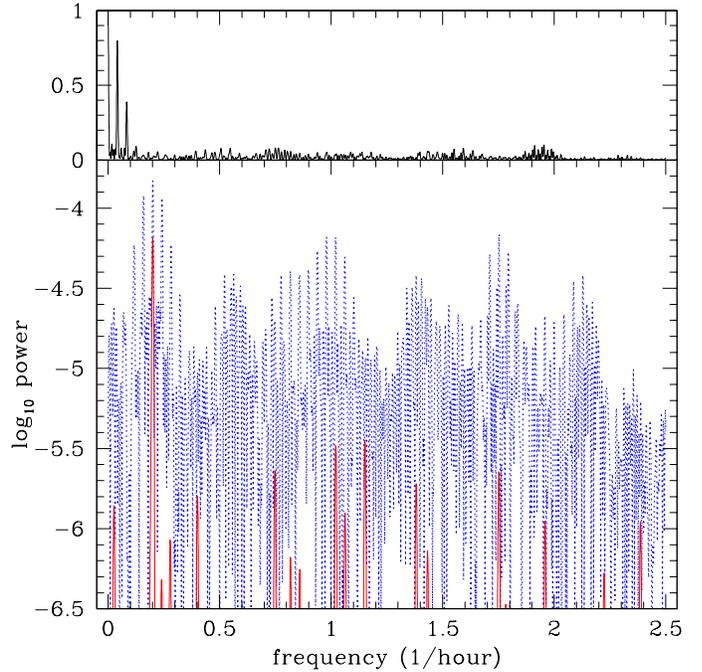,width=0.5\textwidth,angle=0}
\caption{Power spectrum for the simulated light curve shown in
Fig.~\ref{simul01_rot}b. The noise level is $\log_{10}(P) = -6.2$. The
same CLEAN parameters were used here as for the real data of
section~\ref{results}. See caption to Fig.~\ref{2m1145_9901_ps}.}
\label{simul01_ps}
\end{figure}

\begin{figure}
\psfig{figure=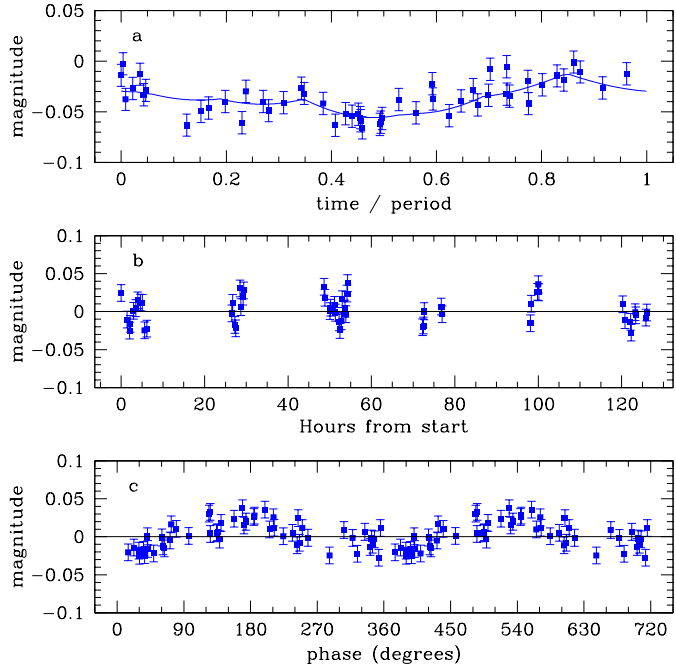,width=0.5\textwidth,angle=0}
\caption{Same as Fig.~\ref{simul01_rot} except now for five dark spots
with random phases. The sampled light curve is again significant
($p<$\,1e-9), and the cleaned power spectrum (Fig.~\ref{simul05_ps})
detects the rotation period at $5.03 \pm 0.10$ hours.}
\label{simul05_rot}
\end{figure}

\begin{figure}
\psfig{figure=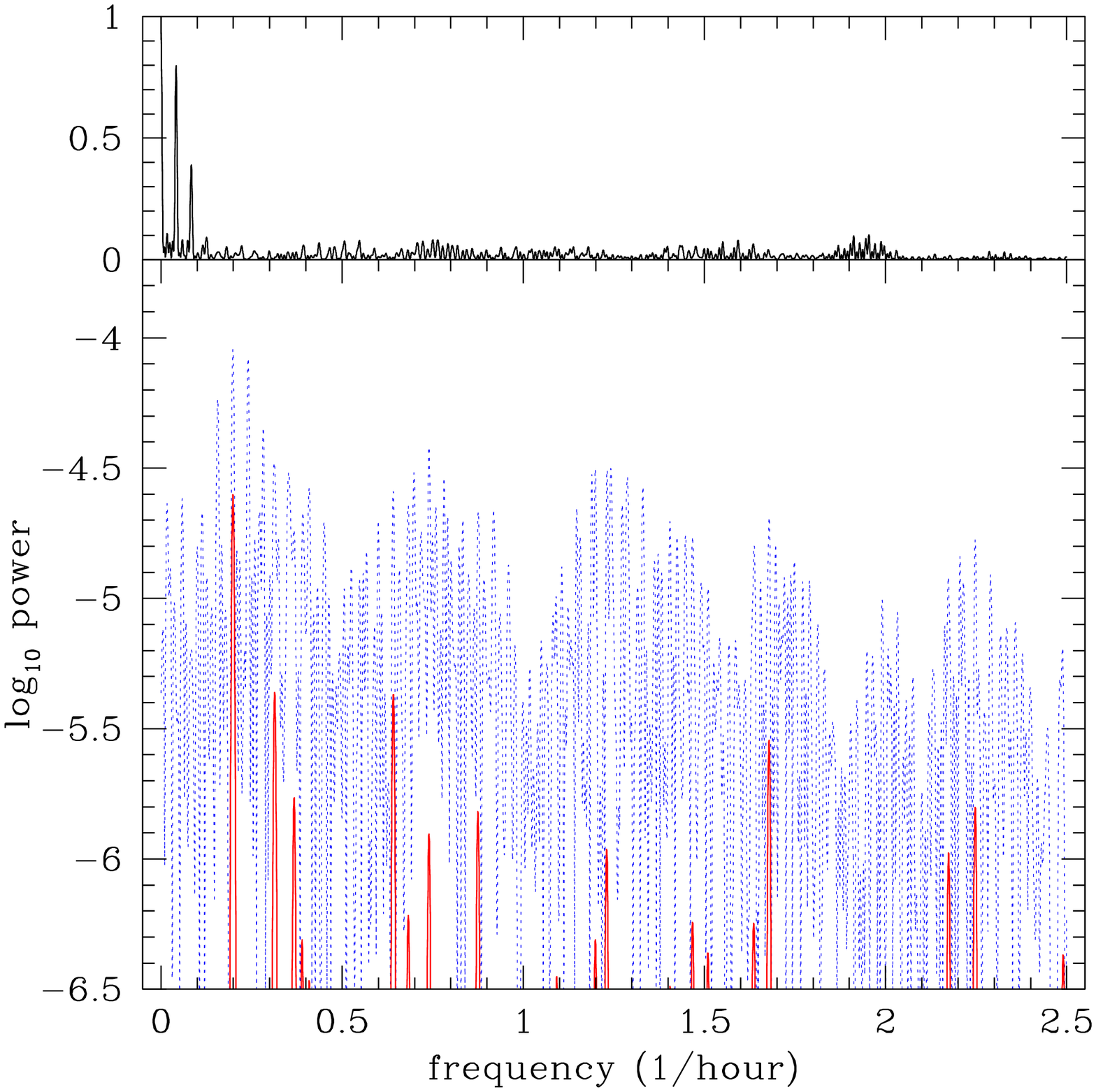,width=0.5\textwidth,angle=0}
\caption{Power spectrum for the simulated light curve shown in
Fig.~\ref{simul05_rot}b. The noise level is $\log_{10}(P) = -6.2$. See
caption to Fig.~\ref{2m1145_9901_ps}.}
\label{simul05_ps}
\end{figure}

We have simulated the appearance of the light curves in a few such
situations.  Fig~\ref{simul01_rot} shows the light curve due to a
single small dark spot on a star which causes a maximum 0.05 magnitude
decrease in brightness.  If we rotate this star with a period of five
hours and observe it with the same noise level and time sampling as
one of our target objects (2M1334) we obtain the power spectrum and
phased light curve in Fig.~\ref{simul01_ps} and
Fig.~\ref{simul01_rot}c respectively. We see that the power spectrum
picks out the rotation period despite the noise and despite the fact
that the light curve is not sinusoidal. Furthermore, the phased light
curve certainly does not resemble a sine wave, yet this {\em is} the
rotation period. Another example is shown in Fig.~\ref{simul05_rot}
where we now have five small dark spots with random longitudes (i.e.\
phases) causing dimmings of 0.011, 0.015, 0.028, 0.030 and 0.034
magnitudes.  Again the star is rotated with a period of five hours and
observed as 2M1334 was. The rotation period is detected by the cleaned
power spectrum (Fig.~\ref{simul05_ps}), yet the phased light curve is
very non-sinusoidal (Fig.~\ref{simul05_rot}c). Note that the power in
the rotation period is reduced compared to the previous simulation.



\begin{figure}
\psfig{figure=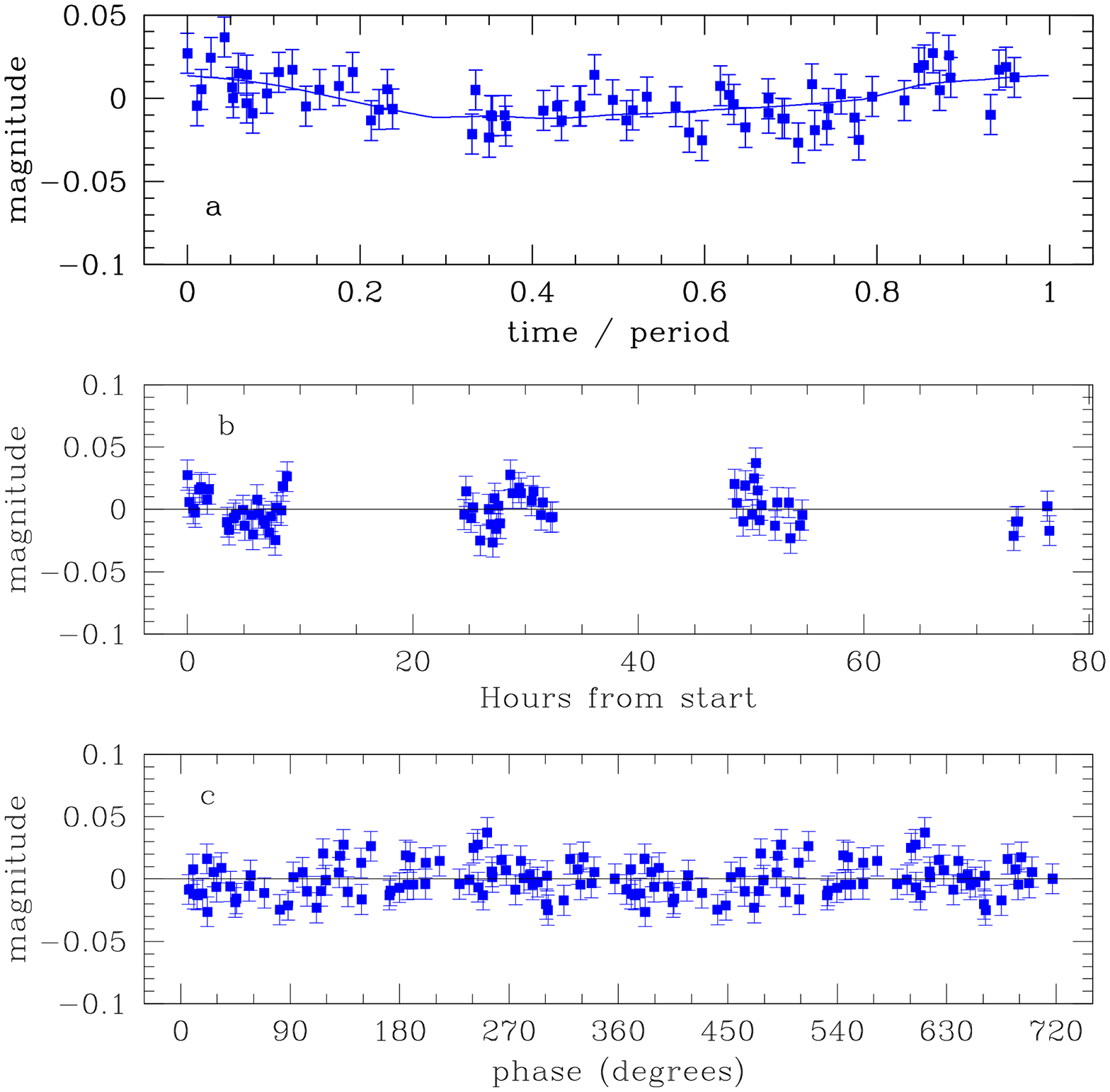,width=0.5\textwidth,angle=0}
\caption{Same as Fig.~\ref{simul01_rot} except now for eight dark and
bright spots with random phases and sampling and noise from
2M1145 (00-02 run). This gives a significant detection, although not
overwhelming ($p=$0.005), yet the cleaned power spectrum
(Fig.~\ref{simul07_ps}) still detects the rotation period of 10
hours.}
\label{simul07_rot}
\end{figure}

\begin{figure}
\psfig{figure=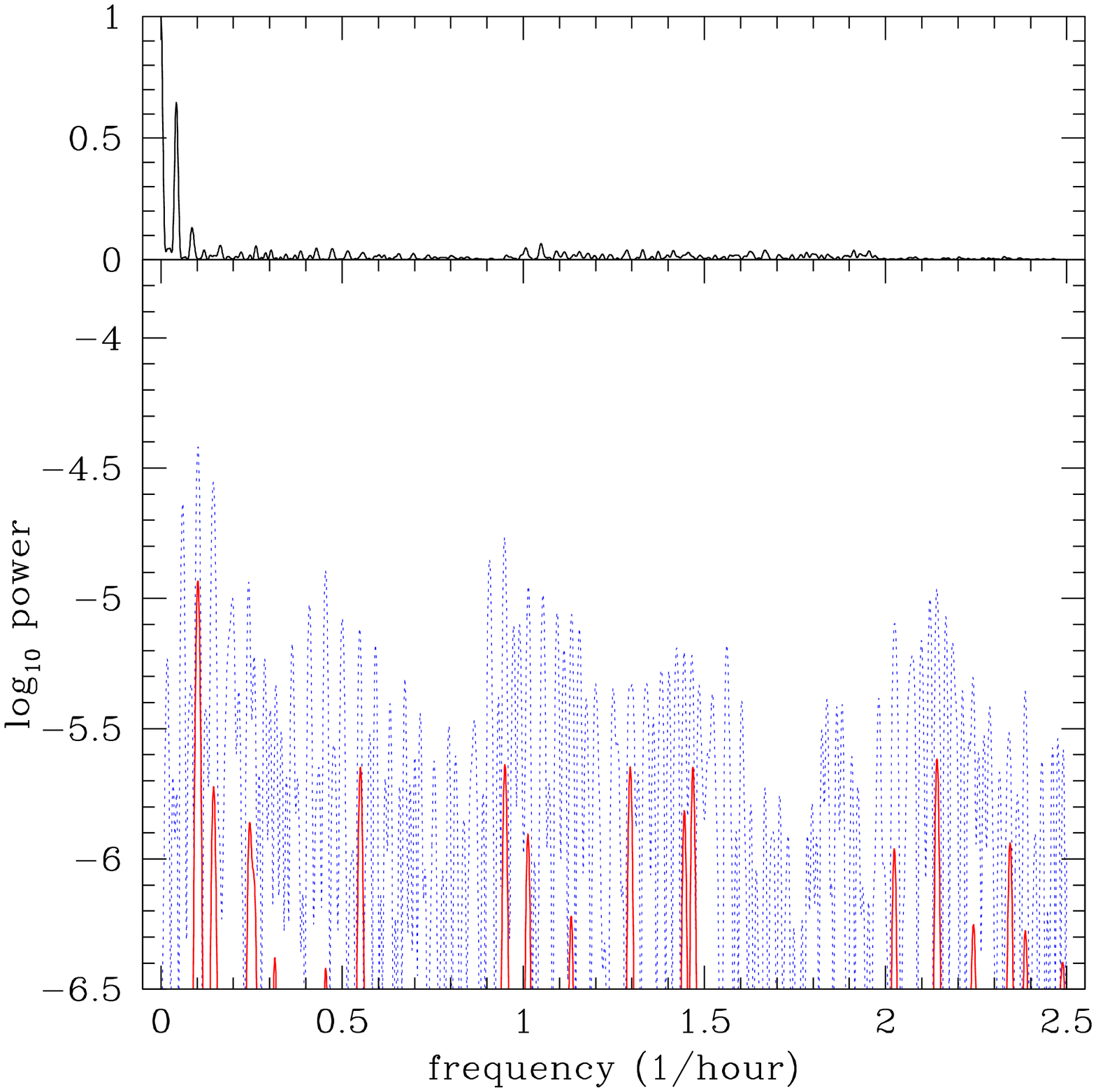,width=0.5\textwidth,angle=0}
\caption{Power spectrum for the simulated light curve shown in Fig.~\ref{simul07_rot}b. The noise level is $\log_{10}(P) = -6.1$. See caption to Fig.~\ref{2m1145_9901_ps}.}
\label{simul07_ps}
\end{figure}

A third simulation is shown in Fig.~\ref{simul07_rot}, which is due to
a star with eight spots rotating with a period of ten hours. Here the
contrast of the individual spots is much smaller, only $-0.008$ to
$+0.014$ magnitudes. The sampling and noise from 2M1145 (00-02 run) is
used and results in a significant variability detection according to
the $\chi^2$ test, but one close to the variable/non-variable cut-off
with $p=$0.005. Despite this low SNR, the rotation
period still clearly stands out in the cleaned power spectrum
(Fig.~\ref{simul07_ps}).

We have carried out many tens of simulations of stars with between one
and ten spots with contrasts between $-0.1$ and $+0.1$ magnitudes and
having random phases, and sampled them using the samping functions of
several objects in this paper.  We found that provided the light curve
showed significant variation (according to our $\chi^2$ criterion)
then the rotation period was always significant ($>$\,10 times the
noise level), and in all but one case was the largest peak.

\subsection{Evidence for the evolution of surface features}

In the light of these simulations, we see that the absence of {\em
sinusoidal} variation in the light curve phased to a certain period
does not rule that out as the rotation period.  Thus the phased light
curve is not a robust means of identifying rotation periods. Moreover,
the absence of any significant peaks seems to imply one of two things:
either the object is not rotating at a period to which we are
sensitive, or the surface features themselves are not stable over the
timescale of observations. A third possibility -- that the contrast of
the spots is too low -- is ruled out because we have already made a
significant detection of variability according to the $\chi^2$
test.\footnote{Another option is that the rotation period is less than
our lower time limit, but this would imply an equatorial rotation
speed in excess of 240\,\kms.}

If the duration of observations is less than a rotation period, the
light curve will show features rotating on and off the limb of the
(unresolved) projected disk: these changes must be represented somehow
in the power spectrum, even though they are not due to long-term
periodicities of the source. If the surface features are not stable,
then the light curve may be even more complex due to the evolution of
individual features. In both cases, we may not expect to see any
dominant periodicities.

Our maximum time span of observations, \tmax, is between 30 and 120
hours, so for us to have observed less than a rotation period, all of
our objects would have to have maximum $v \sin i$ values of between 1
and 4\,\kms\ (assuming a radius of $0.1 R_{\odot}$).  However, this is
inconsistent with the results of Basri et al.\ (\cite{basrietal00}), who
report much higher $v \sin i$ values (10--60\,\kms) for all but one of
a sample of 17 late M and L dwarfs in the field which were not
selected with any known bias for rapid rotation.\footnote{The periods
and rotation velocities can be made consistent if the modulating
features are at a distance of a solar radius from the rotation axis.
However, even the youngest, warmest objects in our sample will
have a radius of no more than $0.2 R_{\odot}$ (Chabrier \& Baraffe
\cite{chabrier00a}).}
Thus our objects probably have rotation periods of order 1 to 10
hours, to which we were certainly sensitive.  Thus the fact that we
have several objects which show no dominant periodicities is
significant, as it appears not to be explainable by rotational
modulation of stable spots. {\em The logical conclusion from the above
arguments is that some of our objects have surface features which
evolve over the duration of our observations.} This applies in
particular to 2M0345, 2M0913, 2M1145 and Calar~3. For 2M1145 we
possibly have more direct evidence of this, as the two light curves
from one year apart show no common periods, despite the fact that
simulations demonstrate we would have detected any likely period due
to stable spots in {\em both} runs. We can imagine that if the
features are themselves changing in brightness then these could
dominate the power spectrum and mask the rotation period. The rotation
period could possibly then be determined through more measurements
over many rotation periods, as the noise level in the power spectrum
would then decrease, whereas the power in the rotation period would
stay constant.

\subsection{Speculation to the physical nature of the surface features}

\begin{figure}
\psfig{figure=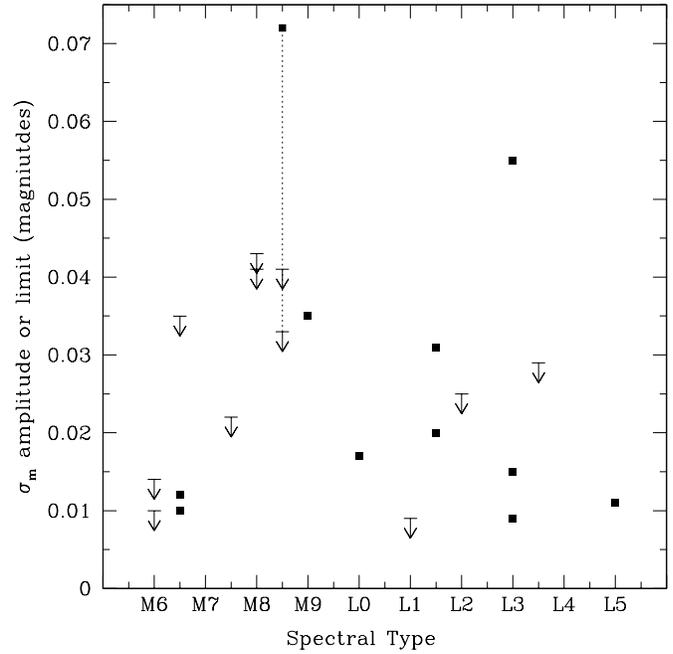,width=0.5\textwidth,angle=0}
\caption{Relationship between variability amplitudes (squares) or upper
limits to variability (arrows) and spectral type. 
S~Ori~45 (M8.5) is plotted as both an amplitude and a limit (connected with
a dotted line) depending on whether the first night of data is
included or not. The plot using $\overline{|m_d|}$ rather than $\sigma_m$
as the amplitude measure is very similar.}
\label{amps}
\end{figure}

Variability in stable stars is often attributed to rotational
modulation of star spots produced by magnetic activity.  In solar-type
stars it is believed to be due to the so-called $\alpha \Omega$
dynamo. This mechanism no longer operates in low mass stars and brown
dwarfs, but as these objects are fully convective, a turbulent dynamo
could come into operation (see Chabrier \& Baraffe \cite{chabrier00a}
and references therein).  M stars often show significant chromospheric
activity, as measured by $\log (L_{\rm H \alpha}/L_{\rm bol})$. Recent
work suggests that this value drops from around $-3.8$ for M7 down to
below $-6$ for L1 and later-type objects (Gizis et
al.~\cite{gizis00}). Basri~(\cite{basri00}) observes a similar decline
and Kirkpatrick et al.\ (\cite{kirkpatrick00}) detected no H$\alpha$
emission (EW $> 2.0$\,\AA) for types later than L4.5.  This suggests
that magnetically-induced surface features may be present on the
surfaces of some of the objects in our sample, but that the contrast
of the spots may decline beyond M7. This is interesting when we
compare it with the relationship between the amplitude of variability
and spectral type, shown in Fig.~\ref{amps}. We see that a larger
fraction of the objects beyond M9 show variability: 7 of 10 equal to
or later than M9, compared to 2 of 9 earlier than M9 (ignoring the
ambiguous detection/non-detection in S~Ori~45).  This is not simply
due a higher detection limit for the earlier type objects, as these
have an average amplitude/detection limit ($\sigma_m$) of 0.025 mag,
compared to 0.023 mag for the later type objects.  If the variability
were due to magnetic spots, we might expect variability to be {\it
less} common among the less active later-type objects, not {\em more}
common as seen here. This trend may be an age effect, as all of our
objects of type M9 and earlier are cluster members with ages less than
120\,Myr. We see no significant relationships between variability
amplitude (or limit) and H$\alpha$ equivalent width.

Another candidate for producing variability is photospheric dust
clouds. It is now well established from detailed modelling of optical
and infrared spectra that late M and L dwarfs have sufficiently cool
atmospheres for solid particles to form (e.g.\ Jones \& Tsuji
\cite{jones97}, Burrows \& Sharp \cite{burrows99}, Lodders
\cite{lodders99}, Chabrier et al.\ \cite{chabrier00}). Whether this
dust stays in suspension in the atmosphere or gravitationally settles
on a short timescale is still an open question.  Basri et al.\
(\cite{basrietal00}) conclude that there must be relatively little
dust opacity on account of the very strong alkali lines in the optical
spectra of L dwarfs.  However, this leaves open the possibility that
dust is present deeper in the photosphere where it would affect the
infrared spectrum.  Models which include dust opacity give better fits
to the near infrared spectra of late M and early L dwarfs than those
which do not (Chabrier et al.\ \cite{chabrier00}). (However, none of
the present models predict accurate near infrared colours for late L
dwarfs, so it appears that the distribution of dust in the atmospheres
of ultra cool dwarfs is more complex than currently appreciated.)
Dust may coalesce into large-scale opaque (dark) clouds, and the
evolution (formation, growth and dissipation) of such clouds over a
few rotation periods could account for our observed variability. These
would have to be relatively large clouds, because many small clouds
evolving independently would have an insignificant net effect on the
light curve. We have seen that ultra cool dwarfs are rapid rotators,
and this (as well as possibly differential rotation) is a likely
driving mechanism for cloud evolution.  These objects are fully
convective, so we can imagine a situation in which dust particles are
convectively cycled up and down in the photosphere.  Dynamical
processes such as turbulent diffusion may well be important for
modelling dust and its formation into clouds, yet such processes are
not taken into account in current atmospheric models. Comparison
with weather patterns seen in solar-system atmospheres must be done
with caution, however, as solar-system planets are significantly
cooler.  This dust cloud explanation appears to be supported by our
observation that variability is more common in later-type (cooler)
objects, i.e.\ those in which more dust can form.

Other options for the variability can be entertained, such as flaring
or outbursts, possibly associated with magnetic activity.  H$\alpha$
flaring is not uncommon in these late-type objects. The very young
objects in $\sigma$~Orionis may still have circumstellar disks from
which they are accreting matter, and variability of the infall (or
even eclipsing by the disk) could account for some variability. There
is, however, no evidence for disks from the infrared observations
(Zapatero Osorio et al.\ \cite{zap00}).  Another possible explanation
is that the variability is due to hotspots from infalling material in
an interacting binary, but this is unlikely to be the explanation in
all cases.

Given the relatively small amounts of data on any one object, it is
difficult to say much about the characteristics of the
variability. However, some of the power spectra are not much different
from random data. If we simulate random light curves from a measured
light curve by reassigning flux measurements to epochs, we find that
the cleaned power spectra often have peaks more than several times the
noise.  While some peaks reported in section~\ref{results} could well
be due to noise, not all peaks can be due to noise when we have a
significant $\chi^2$ detections.  There are several random processes
intrinsic to the star which could produce the observed light curves,
such as the independent evolution of many surface features.  Chaotic
processes can also give the appearance of a random process when
observed in certain parameter spaces.

\section{Summary}

We have presented light curves for 21 late M and L dwarfs to probe
variability on timescales between a fraction of an hour to over 100
hours. 11 objects showed evidence for variability at the 99\%
confidence level according to a $\chi^2$ test, with amplitudes between
0.009 and 0.055 magnitudes (RMS).  Of these objects, four (2M1145,
2M1334, SDSS~0539, S~Ori~31) showed strong evidence for variability
(confidence greater than 99.99\%).  It has been shown how a careful
data reduction and analysis of the errors ensures the reliability of
this test.  The ten non-detections have upper limits on their
RMS amplitudes of between 0.009 and 0.043 magnitudes.

A power spectral analysis was performed on all variable objects using
the CLEAN algorithm.  In a few cases (2M1146, 2M1334, SDSS~0539,
S~Ori~31) there were significant periodicities (at $5.1 \pm 0.1$,
$2.68 \pm 0.13$, $13.3 \pm 1.2$ and $7.5 \pm 0.6$ hours respectively)
which dominated the power spectra.  For 2M1334, SDSS~0529 and S~Ori~31
these may be the rotation periods. We demonstrated with simulations
that the rotation period does not necessarily produce {\em sinusoidal}
variation in the light curve: Thus these periods can only be confirmed
or refuted with longer-term monitoring with more complete coverage.
The 5.1 hour period for 2M1146 was shown not to be the rotation period
on the basis of an inconsistency with the $v \sin i$ measurement of
Basri et al.  The remaining seven significantly variable light curves
did not show dominant periods, and in at least three cases (2M0345,
2M0913, Calar~3) there are not even any significant periods.  Our
simulations showed that we would have detected any plausible rotation
periods for these objects based on $v \sin i$ measurements.  We
therefore concluded that the lack of significant periods was due to
the evolution of the features on timescales shorter than our
observation span, and that these ``wash out'' the rotation period in
the power spectrum. 2M1145 showed no common periodicities in two
separate significantly variable light curves obtained on year apart,
thus supporting this view.

We found that variability is more common in objects later than M9: 7
of 9 objects later than M9 are variable, compared to only 2 of 9
earlier.  This may be related to the observation of Gizis et al.\ that
chromospheric activity declines significantly from M7 to L1, and
perhaps points to the variability in the late-type objects having a
non-magnetic origin; photospheric dust clouds were highlighted as a
likely cause. Gaining more insight into the nature of the variability
observed in this paper will be the next challenge.

\section*{Acknowledgements}
CBJ is very grateful to Harry Lehto for use of his {\sc CLEAN} code and
information on its application and interpretation.  CBJ also thanks
Bill Herbst for useful discussions and an independent analysis of one
of the light curves, and Pablo Cincotta for use of his phase
dispersion minimisation code. The authors thank Barrie Jones for
comments on a draft manuscript. This work is based on observations made
with the 2.2m telescope at the German--Spanish Astronomical Center at
Calar Alto in Spain.


\begin{thebibliography}{99}
\bibliographystyle{unsrt}
  
\bibitem[1999]{paper1}Bailer-Jones C.A.L., Mundt R., 1999, A\&A 348, 800 (paper I)

\bibitem[2000]{basri00}Basri G., 2000, ASP Conf.\ Ser., in press

\bibitem[2000]{basrietal00}Basri G., Mohanty S., Allard F., et al., 2000, ApJ 538, 363

\bibitem[1999]{burrows99}Burrows A., Sharp C.M., 1999, ApJ 512, 843

\bibitem[1999]{bejar99}
B\'ejar V.J.S., Zapatero Osorio M.R., Rebolo R., 1999, ApJ 521, 671

\bibitem[2000]{chabrier00a}
Chabrier G., Baraffe I., 2000, ARA\&A 33, 337

\bibitem[2000]{chabrier00}
Chabrier G., Baraffe I., Allard F., Hauschildt P., 2000, ApJ 542, 464

\bibitem[1995]{cincotta95}
Cincotta P.M., M\'endez M., N\'e\~nez J.A., 1995, ApJ 449, 213
 
\bibitem[1975]{deeming75}
Deeming T.J., 1975, Astrophysics \& Space Science 36, 137

\bibitem[1978]{driscoll}Driscoll W.G.\ (ed.), {\em Handbook of Optics}, Optical Society of
	America/McGraw-Hill, 1978, p. 14-59.

\bibitem[2000]{fan00}
Fan X., Knapp G.R., Strauss M.R., et al., 2000, AJ 119, 928

\bibitem[2000]{gizis00}
Gizis J.E., Monet D.G., Reid I.N., Kirkpatrick J.D., Liebert J., Williams R.J., 2000, AJ 120, 1085

\bibitem[2001]{herbst01}Herbst W., Bailer-Jones C.A.L., Mundt R., Meisenheimer K., 
	Wolf C., Wackerman R., 2001, in preparation

\bibitem[1997]{jones97}Jones H.R.A., Tsuji T., 1997, ApJ 480, L39

\bibitem[1999]{kirk99}
Kirkpatrick J.D., Reid I.N., Liebert J., et al., 
	1999, ApJ 519, 802

\bibitem[2000]{kirkpatrick00}
Kirkpatrick J.D., Reid I.N., Liebert J., et al., 
	2000, AJ 120, 447

\bibitem[1999]{koerner99}
Koerner D.W., Kirkpatrick J.D., McElwain M.W., Bonaventura N.R., 1999, ApJ 526, L25

\bibitem[1999]{lodders99}Lodders K., 1999, ApJ 519, 793

\bibitem[1998]{martin98}Mart\'\i n E.L., Basri G., Zapatero-Osorio M.R., Rebolo R., Garc\'\i a L\'opez R.J.,
	1998, ApJ 507, 41

\bibitem[2000]{nakajima00}Nakajima T., Tsuji T., Maihara T., et al., 
	2000, PASJ 52, 87

\bibitem[1995]{rebolo95}Rebolo R., Zapatero Osorio M.R., Mart\'\i n E.L.,
	1995, Nat 377, 129

\bibitem[2000]{reid00}Reid I.N., Kirkpatrick J.D., Gizis J.E., et al., 2000, AJ 119, 369

\bibitem[1996]{rebolo96}Rebolo R., Mart\'\i n E.L., Basri G., Marcy G.W., 
	Zapatero-Osorio M.R.,1996, ApJ 469, L53

\bibitem[1987]{roberts87}
Roberts D.H., L\'ehar J., Dreher J.W., 1987, AJ 93, 968

\bibitem[1999]{terndrup99}Terndrup D.M., Krishnamurthi A., Pinsonneault M.H., Stauffer J.R., 1999, ApJ 118, 1814

\bibitem[1999]{tinney99}Tinney C.G., Tolley A.J., 
	1999, MNRAS 304, 119

\bibitem[1991]{young91}Young A.T., Genet R.M., Boyd L.J., et al., 1991, PASP 103, 221

\bibitem[1999]{zap99}Zapatero Osorio M.R., Rebolo R., Mart\'\i n E.L., et al.,
	1999, A\&AS 134, 537

\bibitem[2000]{zap00} Zapatero Osorio M.R., B\'ejar V.J.S., Mart\'\i n E.L., Rebolo R.,
	Barrado y Navascu\'es D., Bailer-Jones C.A.L., Mundt R., 2000, Sci 290, 103


\end{thebibliography}
\end{document}